\documentclass[draft]{agujournal2019}
\usepackage{url} 
\usepackage{lineno}
\usepackage[finalnew]{trackchanges} 
\usepackage{soul}

\usepackage{graphicx, amsmath, array, algorithmic}
\usepackage{subfig, gensymb}
\usepackage{soul}

\addeditor{AH}

\graphicspath{{./}{figures/}}

\draftfalse

\journalname{Radio Science}

\begin{document}

%
%


\title{Measuring the Earth's Synchrotron Emission from Radiation Belts with a Lunar Near Side Radio Array}

%
%




\authors{Alexander Hegedus$^1$, Quentin Nenon$^2$, Antoine Brunet$^3$, Justin Kasper$^1$, Ang\'{e}lica Sicard$^3$, Baptiste Cecconi$^4$, Robert MacDowall$^5$, Daniel Baker$^6$}


\affiliation{1}{University of Michigan, Department of Climate and Space Sciences and Engineering, Ann Arbor, Michigan, USA}
\affiliation{2}{Space Sciences Laboratory, University of California, Berkeley, CA, USA}
\affiliation{3}{ONERA / DPHY, Universit\'{e} de Toulouse, F-31055 Toulouse – France} 
\affiliation{4}{LESIA, Observatoire de Paris, Université PSL, CNRS, Sorbonne Université, Univ. de Paris, Meudon, France}
\affiliation{5}{NASA Goddard Space Flight Center, Greenbelt, MD, USA}
\affiliation{6}{University of Colorado Boulder, Laboratory for Atmospheric and Space Physics, Boulder, Colorado, USA}




\correspondingauthor{Alexander Hegedus}{alexhege@umich.edu}




\begin{keypoints}
\item Synchrotron emission between 500-1000 kHz has a total flux density of 1.4-2 Jy at lunar distances
\item\add[AH]{A 10 km radio array with 16000 elements could detect the emission in 12-24 hours with moderate noise} \remove[AH]{A 10 km radio array with O(16000) elements could measure the belts 1-2 times/day with moderate noise}
\item Changing electron density can make detections 10x faster at lunar night, 10x slower at lunar noon 
\end{keypoints}

%
%

%
%


\begin{abstract}
The high kinetic energy electrons that populate the Earth's radiation belts emit synchrotron emissions because of their interaction with the planetary magnetic field.  A lunar near side array would be uniquely positioned to image this emission and provide a near real time measure of how the Earth's radiation belts are responding to the current solar input.  The Salammb\^{o} code is a physical model of the dynamics of the three-dimensional phase-space electron densities in the radiation belts, allowing the prediction of 1 keV to 100 MeV electron distributions trapped in the belts.  This information is put into a synchrotron emission simulator which provides the brightness distribution of the emission up to 1 MHz from a given observation point.  Using Digital Elevation Models from Lunar Reconnaissance Orbiter (LRO) Lunar Orbiter Laser Altimeter (LOLA) data, we select a set of locations near the Lunar sub-Earth point with minimum elevation variation over various sized patches where we simulate radio receivers to create a synthetic aperture.  We consider all realistic noise sources in the low frequency regime.  We then use a custom CASA code to image and process the data from our defined array, using SPICE to align the lunar coordinates with the Earth.  We find that for a moderate lunar surface electron density of 250/cm$^3$, the radiation belts may be detected every 12-24 hours with a 16384 element array over a 10 km diameter circle.  Changing electron density can make measurements 10x faster at lunar night, and 10x slower at lunar noon.

\end{abstract}

\section*{Plain Language Summary}
The Earth's Ionosphere is home to a large population of energetic electrons that live in the balance of many factors including input from the Solar wind, and the influence of the Earth's magnetic field.  These energetic electrons emit radio waves as they traverse Earth's magnetosphere, leading to short-lived, strong radio emissions from local regions, as well as persistent weaker emissions that act as a global signature of the population breakdown of all the energetic electrons.  Characterizing this weaker emission (Synchrotron Emission) would lead to a greater understanding of the energetic electron populations on a day to day level.  A radio array on the near side of the Moon would always be facing the Earth, and would well suited for measuring its low frequency radio emissions.  In this work we simulate such a radio array on the lunar near side, to image this weaker synchrotron emission.  The specific geometry and location of the test array were made using the most recent lunar maps made by the Lunar Reconnaissance Orbiter.  This array would give us unprecedented day to day knowledge of the electron environment around our planet, providing reports of Earth's strong and weak radio emissions, giving both local and global information.


\section{Introduction} \label{sec:intro}

Understanding the energetic electron environment below 6 Earth radii has long been an area of scientific interest as well as practical concern.  This information helps us to understand the radiation dosages that spacecraft at different orbits are likely to see over time, which in turn goes into the Total Ionizing Dose (TID) the spacecraft is designed to be tolerant to.  The response of the radiation belts to solar input can elicit a variety of responses, complicating the calculation of how much radiation a given spacecraft has actually been exposed to so far.  In order for spacecraft industries to track the predicted remaining lifetimes of all their satellites, it would be useful to have some real measure of how many energetic electrons were in Earth's radiation belts at any given time.  This is especially useful for the many satellites that do not have energetic particle detectors to measure their received radiation dose.  Even with detectors, existing satellites can give only single point \textit{in situ} measurements of the electron distribution from a stable orbit.  Measurements of the global synchrotron emission could yield a view of the bigger picture by providing a proxy measurement of the global electron distribution, providing useful constraints for space weather forecasting models and TID calculations.  An array capable of such measurements would also be able to localize auroral transient events with high precision, providing local, small scale electron data in addition to global data. 

Many planets with magnetic fields have radiation belts from trapped electrons to some degree.  However, Jupiter is the only outer planet that has had synchrotron emission detected from its radiation belts, making it a good case to look at in order to understand what to expect in observing the Earth's synchrotron emission.  Jupiter's strong magnetic field traps high energy electrons up to 10s of MeV \cite{Bolton2002}, and these stable energetic electron belts produce synchrotron emission in the decimeter (DIM) wavelength range \cite{Carr83}.  The physics of synchrotron emission are well understood at this point \cite{Pacho70}: an electron at a certain energy will release photons at a broad spectrum of frequencies corresponding to the the envelope of the summation of harmonics of the cyclotron frequency.  The cyclotron frequency $f_c$ is the frequency in Hz at which a charged particle such as an electron with mass $m$ and charge $q$  gyrates around a magnetic field with field strength $B$ in Gauss (G).  

\begin{align}
  f_{c} = \frac{qB}{2\pi m}
  \label{eqn_cyclo}
\end{align}

An electron with energy $E$ in Mega electronvolts (MeV) and pitch angle $\alpha$ will emit a broad range of frequencies corresponding to the envelope of the summation of cyclotron harmonics, with a maximum at around \add[AH]{$f_{peak}$} \remove[AH]{$f_{max}$} MHz, where

\begin{align}
  f_{peak} \approx 4.8 E^2 B \sin \alpha  
  \label{eqn_synch}
\end{align}

It is important to note that \add[AH]{$f_{peak}$} \remove[AH]{$f_{max}$} is the frequency at which the maximum amount of photons are being emitted, not the highest frequency with any emission.  

The energy of the Jovian radiation belt electrons that contribute to the DIM emission typically ranges from hundreds of keV (i.e., barely relativistic electrons) to several hundred MeV (i.e., ultra-relativistic electrons).  It is generally accepted that at Jupiter this synchrotron emission from high-energy electrons dominates at frequencies 100-3000 MHz, while thermal emission overtakes it at higher frequencies.  This synchrotron emission is characterized by large angular extent relative to the visible disk and by its high degree of linear polarization. 

With the basic physics of synchrotron emission pinned down, a challenge in recent years was to deduce the spatial and energy distribution of electrons to allow to best reproduction of the observed 2D and 3D maps of radio emission \cite{Costa08, Girard:2016gy}.  This has been achieved with synthetic 2D radio maps that have excellent agreement with radio observations  \citeA{Costa01, Sicard04, Nenon2017}.  These results used a version of the Salammb\^{o} code tuned to Jupiter's environment to model the physics in the radiation belt emissions \cite{Salammbo95} \cite{Salammbo96} \cite{Salammbo2000}.  

Observation of the Jovian radiation belt synchrotron emissions has enabled major progress in the understanding of the radiation belts physics and average distribution (\citeA{Nenon2017} and references therein). They also enabled the study of short time scale changes (hours to months) in the electron distributions near Jupiter related to cometary impacts \cite{Costa11} or to the solar wind \cite{Costa14}. Long time scale dynamics (years) linked to the solar wind have also been revealed \cite{Han18}.

Earth's radiation belts also have keV and MeV electrons as confirmed by \cite{Pierrard2019} using the EPT (Energetic Particle Telescope) onboard the satellite PROBAV, as well as measurements from THEMIS \cite{themis08}.  These energetic electrons should also produce synchrotron emission, the brightness of which reveals the electron distribution across different energy levels.  In theory, one could use measurements from an array with sufficient sensitivity to measure the brightness spectrum in small bandwidths from 1 MHz and below, and back out a detailed proxy for the current global electron energy distribution.  In reality, signal to noise concerns mean that for initial arrays, large bandwidths will have to be combined in order to make good detections.  Even with large bandwidths, this would still be valuable information for understanding the global response of the Earth's radiation belts to space weather.  In this work we design an initial array that could do some baseline imaging of the radiation belts from the lunar sub-Earth point.

An outline of the paper is as follows: in Section 2, we describe how we use the Salammb\^{o} code to simulate the Earth's electron environment from which we extract the resulting synchrotron emission as seen from the lunar surface.  These simulations will be used as ground truth images.  Ground truth images are representations of the target the array will image.  These are input into array simulations and compared to the output images to evaluate the array's performance in capturing the details from the input.   In Section 3, we outline all competing noise sources in this observing frequency range and decide on an operational science bandwidth.  In Section 4 we design a pathfinder array on the lunar surface that can detect and image Earth's synchrotron emission.  In Section 5, the results of our simulations are discussed.  In Section 6 we outline future work to be done and future related missions.

\section{Generating Ground Truth Images}

As seen in Equation \ref{eqn_synch}, the peak emission frequency for a given electron energy level is proportional to electron energy $E^2$ and $B$, the strength of the planetary magnetic field.  The magnetic moment of Jupiter is\add[AH]{ $1.59 \cdot 10^{30} G/cm^3$}, while Earth's is $ 2.10\cdot 10^{25} G/cm^3$ \cite{Jun05}.  Jupiter has a peak flux of $\geq$ 1 MeV electrons of $10^8$ electrons/$cm^2/s$ while Earth has a peak flux of $\geq$ 1 MeV electrons of $10^7$ electrons/$cm^2/s$.  The most energetic electrons in Earth's magnetosphere at 6 Earth radii are below 10 MeV, while the most energetic electrons in Jupiter's magnetosphere at 9.5 Jovian radii are above 1000 MeV \cite[Fig. 3]{Jun05}.  This implies that the expected emission at Earth will be at a far lower frequency than seen at Jupiter.  It is partially for this reason that progress on imaging the Earth's radiation belts has been significantly slower than for those of Jupiter, since there is not a straightforward way to image the global structure of the belts when you are trying to do it from a small portion of the globe itself.  There is also the issue of the ionospheric cutoff, which precludes radio waves below 10 MHz from making all the way through the ionosphere to the Earth's surface.  This means that 1 MHz signals generated near the topside ionosphere could not make it down to the ground for detection.    

A lunar near side array would be uniquely positioned to measure the belts, and provide a near real time measure of how the Earth's radiation belts are responding to the current solar input.  The Salammb\^{o} code solves the three-dimensional phase-space diffusion equation while modeling Coulomb collisions with neutral and plasma populations around Earth, wave-particle interactions, radial diffusion and magnetopause shadowing induced dropouts.  It models the radiation belts in a computational domain that extends from L=1 to L=10 and uses the International Geomagnetic Reference Field (IGRF) decentered tilted dipole magnetic field model. The simulation starts 50 days before the two target dates with empty radiation belts. At L=10, the modern iteration of the Salammb\^{o} code uses the Time History of Events and Macroscale Interactions during Substorms/Solid State Telescope (THEMIS-SST) data set of electron distributions up to several hundred keV as an outer boundary condition \cite{Maget15}.  The SST instruments aboard THEMIS provide measurements of omnidirectional electron flux in 11 energy channels ranging from 31 keV to 720 keV, as well as unidirectional ones resolving eight pitch angles between 0$\degree$ and 180$\degree$ \cite{themis08}.  The model also takes Kp as an input, which parameterizes radial diffusion strength and plasmapause position.  An Ensemble Kalman Filter (EnKF) is employed by the model for data assimilation, leading to improvements in the predictions.  The output is a global model of the trapped electrons in the radiation belts from 1 keV to 100 MeV.  

\begin{figure*}[t]
\centering
\includegraphics[width=\textwidth]{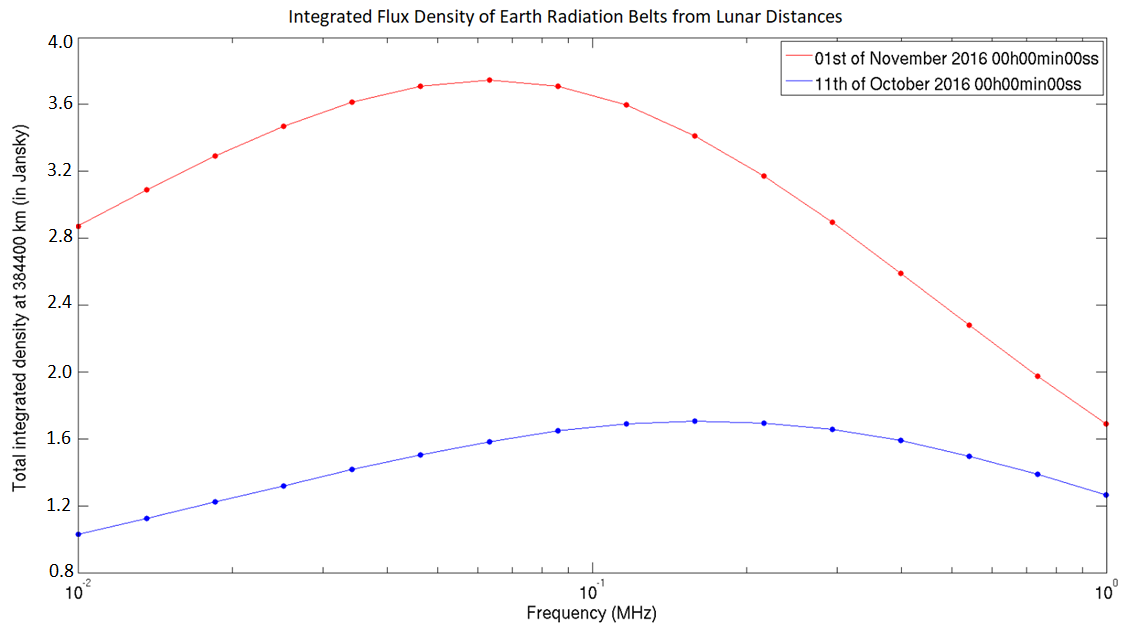}
\caption{\textbf{Integrated spectral flux density of synchrotron radiation from lunar orbit.  The red line is the modeled synchrotron flux density spectrum for a stormy period on November 1st 2016 when electron fluxes were higher as a result of the impact of solar wind structures onto the Earth's magnetosphere.  The blue line is the modeled synchrotron spectrum for a calmer period from October 11th 2016 when the electron flux was lower. Both spectra are scaled for an observer on the lunar surface.} }
\label{fig_brightness}
\end{figure*}

We ran two simulations on a modern version of the Salammb\^{o}-EnKF code, a ``quiet time'' which represents what can be seen on 11th of October 2016, and a ``storm time'' on 1st of November 2016 when electron fluxes were higher as a result of the impact of solar wind structures onto the Earth's magnetosphere. Only these two dates are used as a research target in this study.  A thorough investigation of the synchrotron radiation emitted by the radiation belts in more extreme configurations, identifying the lowest and highest possible electron fluxes, is left for future work, as are the time dynamics and response of synchrotron radiation to solar wind events.  The output of these two simulated periods are then analyzed to provide realistic predictions of the brightness of the synchrotron emission up to 1 MHz.  To do so, the synchrotron emission simulator developed at ONERA for Jupiter and Saturn has been adapted to Earth (for details on the synchrotron simulator, see \cite{Nenon2017} and references therein).  The synchrotron emission simulator takes the electron distribution in the belts as input, as well as the magnetic field of the planet and the position of the observer.  The output is a 2D image of the total intensity (first Stokes parameter) of the synchrotron emissions for a given frequency. It is expressed as brightness temperatures (in Kelvin) and can be converted to Jansky/beam or Jansky/pixel.  For lunar distances, the output images are 400x400 \add[AH]{2.28 arcminute} pixels, for a total area in the sky of 15.2 degrees.  On average, the angular size of Earth from the lunar surface is 1.91 degrees, so 1 Earth radius is about 25 pixels in this scale.

\begin{figure*}[h!]
\centering

\includegraphics[width=1.0\textwidth]{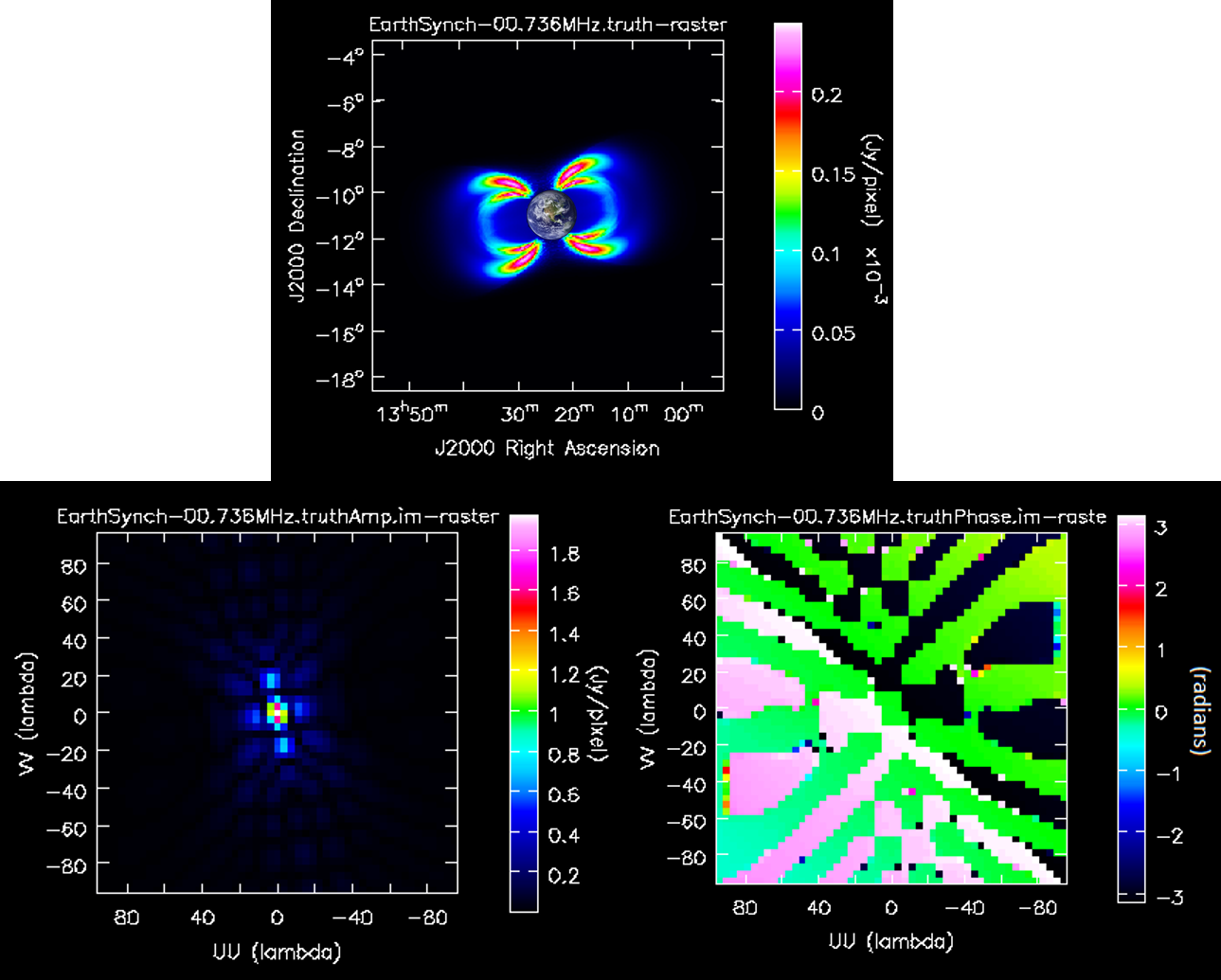}
\caption{\textbf{Simulated Radiation Belt Emission \& Fourier Transform. \textit{Top}: Truth image of synchrotron emission from radiation belts at Lunar Distances.  This is what goes into the simulated array pipeline and is compared to the output.  Brightness map created from Salammb\^{o} electron simulation data.  The 1.91$\degree$ Earth is added in for a scale indicator.\textit{Left}: 2D Fourier Transform Amplitude. \textit{Right}: 2D Fourier Transform Phase (radians). }}
\label{fig_truth}
\end{figure*}

We generated brightness maps from 0.1 to 1 MHz scaled to lunar distances with an overall spectral flux density in the 1 - 3.75 Jy range.  These spectral flux density totals at lunar distances are seen in Figure \ref{fig_brightness}.  An example of the brightness map for a stormy period at 736 kHz is seen in Figure \ref{fig_truth} (a).  The other parts of Figure \ref{fig_truth} show the 2D Fourier Transform of the sky brightness pattern, which is what the synthetic aperture described in Section 4 will be sampling.  One should note that the synchrotron intensities are directly proportional to the flux of trapped electrons at a given energy, and a variation of a factor of 10 is easily encountered in the Earth radiation belts during extreme solar wind events.

Brightness maps of Earth synchrotron emissions can exhibit and confirm what has been observed by the Van Allen Probes, that found an ``impenetrable'' barrier at L=2.8, below which energetic electrons cannot penetrate \cite{Baker2014}.  This barrier has been observed over the course of many years \cite{Baker19}, and is thought to originate from a magnetically confined bubble of very low frequency (VLF) wave emissions of human origin \cite{Foster16}.  Equation \ref{eqn_synch} implies that at the highest synchrotron frequencies, most the contribution comes from the highest energy electrons at the strongest magnetic field strengths.  This emission, seen in Figure \ref{fig_truth}a, maps the synchrotron brightness at 734 kHz, on the high end of Earth's synchrotron emission.  As expected, brightest emission is near the footpoints of the magnetic L shells where the magnetic field is stronger, and there is almost no emission below the barrier of L=2.8 at these highest frequencies, implying a lack of energetic electrons in agreement with observations from the Van Allen Probes.

\section{Noise}

We follow \citeA{Zaslavsky11} which gives the equations needed for calibrating the response of a short dipole antenna.  They use these equations to do the antenna calibration of the STEREO/WAVES (S/WAVES) radio instrument \cite{Bale2008} onboard the STEREO spacecraft \cite{Bougeret2008}, using the Galactic radio background as a reference source.  They considered 3 main sources: amplifier noise, quasithermal noise from free electrons, and Galactic background radiation from the Milky Way.  Out study will include all relevant potential sources of noise for measurements requiring sensitivity on the order of 1 Jy.  These competing sources can be put into 3 classes of signals: removable constants, transients, and unavoidable noise.

\subsection{Removable Constant Background Radiation }
These noise sources are static in nature and must be understood to the sub-1 Jy level in order to remove them and detect the synchrotron emission from Earth's radiation belts.

\subsubsection{Galactic Background Radiation}

Galactic Thermal Noise from the Milky Way has been characterized extensively before \cite{Cane79} \cite{Novaco78}.  The model from \cite{Novaco78} is seen in Figure \ref{fig_noiseLevels} alongside other large noise sources.  From spinning antenna experiments, it's been thought that the Galactic brightness below 10 MHz is mostly isotropic.  Modulations as a function of the observed solid angle are around 20\% at 0.3 MHz, and decreases down to near 0\% at 3.6 MHz \cite{manning01}, with the Galactic poles having a slight brightness enhancement.  

This implies for nonzero baselines, there is a maximum power of 20\% of the average Galactic brightness. In order to detect the radiation belts, this is a foreground source that will needed to be understood to around a $10^{-5}$ level in order to not confuse it with the weaker synchrotron emission.  This will be a mapping effort that has happened at higher frequencies, but never to such a degree for the lowest frequency radio sky.  Because Galactic background radiation is the largest static source in the low frequency sky, it is also the most useful for calibration of the antennas \cite{Zaslavsky11}.

\subsubsection{Blackbody Noise}
There are 3 main blackbody Sources to consider: the Earth, the Sun, and the lunar surface itself.  Earth has an equivalent blackbody temperature of 288 K.  Following Plank's law \cite{Planck14}, a maximum blackbody brightness is found to be 8.8$\cdot10^{-26}$ W/m$^2$/sr/Hz at 1 MHz, and decreases for lower frequencies.  One can multiply these values by $4 \pi \cdot (\frac{R_E}{D_{EM}})^2$ to account for the inverse square decrease in intensity from the Earth's surface to the Moon and convert to spectral flux density units W/m$^2$/Hz to make a Jansky comparison.  This decreases the total signal from the Earth's blackbody output to an integrated 30.4 mJy for 1 MHz at lunar distances and is less strong at lower frequencies.  This effect is small and fairly constant and may be subtracted out of the data on a per channel basis.  In the scale of our truth images, this 30.4 mJy signal is spread throughout the 1960 or so pixels that make up the Earth, giving an average of about 0.0156 mJy/pix.  This over an order of magnitude below the peak mJy/pix values for the radiation belt, and is ignored in our simulations.  

\begin{table*}[t]
\center

\caption{
\textbf{ Characteristics of constant sources as seen from a lunar based radio array} } 

\begin{tabular}{|c|c|c|}
\hline
    \shortstack{Constant Source} &  \shortstack{Lunar Flux \\Density 1 MHz} & Notes \\
    \hline
    \shortstack{Galactic \\ Brightness} &  $5 \cdot 10^{6}$ Jy &  Acts like correlated noise\\
    \hline
    \shortstack{Earth \\ Blackbody} &$3 \cdot 10^{-2}$ Jy & \shortstack{$1.9\degree$ circle from the Moon \\ 273 K } \\
    \hline
    \shortstack{Lunar \\ Blackbody} & \shortstack{38 - 144 Jy \\ Night to Day} & \shortstack{Added to background noise \\ 100 - 373 K } \\
    \hline
    \shortstack{Solar \\ Blackbody} &$4.84 \cdot 10^{-2}$ Jy & \shortstack{30 armin circle from the Moon \\ 5800 K } \\
    \hline
    \shortstack{Coronal \\ Thermal Bremsstrahlung} & $>$ 100s of Jy & \shortstack{Variable Morphology \\ Several solar radii across } \\
    \hline
\end{tabular}

 \label{table_constants}

\end{table*}

The Moon has an average black body temperature of 271 K, but can have temperatures of 373 K in the daytime (yielding a 1 MHz blackbody noise of 1.14$\cdot10^{-25}$ W/m$^2$/sr/Hz or 1.44$\cdot10^{-24}$ W/m$^2$/Hz = 144 Jy) and 100 K at night (yielding a 1 MHz blackbody noise of 3.06$\cdot10^{-26}$ W/m$^2$/sr/Hz or 3.85$\cdot10^{-25}$ W/m$^2$/Hz = 38.5 Jy).  Since this is from the surface of the Moon itself, and not from a small area in the sky, this blackbody noise will add random thermal noise to our system, but is less than 3 orders of magnitude below other noise sources even in optimistic amplifier limited noise regimes.  We will therefore not include it in our simulations.  A summary of the basic characteristics of these constant background radiation noise sources can be seen in Table \ref{table_constants}.

The Sun has a blackbody temperature of 5800 K, giving a maximum surface brightness of 1.78$\cdot10^{-24}$ W/m$^2$/sr/Hz at 1 MHz.  The mean radius of Sun is 696,000 kilometers and 1 AU is 1.496$\cdot 10^{8}$ kilometers.  Multiplying again by $(\frac{r_1}{r_2})^2 \cdot 4 \pi$ yields 4.838$\cdot10^{-28}$ W/m$^2$/Hz or 48 mJy for the flux density at the Moon. This originates from a \add[AH]{12 arcminute} circular source, and would correspond to to about 2.5 mJy/pixel when spread through the 20 or so pixels the sun would take up in the resolution of our truth images.  These levels are similar to those in the signal in the Earth's synchrotron emission at lunar distances, and will thus have to be removed in post processing with CLEAN \cite{hogbom74} or a similar algorithm if it is close to the Earth in the sky.  A more advanced multiscale method like MultiScale-CLEAN (MS-CLEAN) \cite{Cornwell2008} may also be used to remove the Sun from the image, using the known size of the Sun as an input to facilitate an direct removal of that sized feature.  Peeling methods \cite{Noordam04} may also be used to remove the influence from this known source in the visibility domain, before the imaging process.  

\add[AH]{Imaging studies of the Sun by the LOFAR array have shown that at lower frequencies, thermal bremsstrahlung emission from the hot 1-2 MK solar corona far outstrips that of the Sun's blackbody emission} \cite{Vocks18}.  \add[AH]{These studies have revealed that as the Sun is imaged in progressively lower frequencies, more of the corona is seen to be emitting.  This emitting region will be several solar radii across in the frequencies shared by Earth's synchrotron emission.  The exact brightness and morphology of this coronal emission is variable and dependent on solar activity.  Similar to mitigating the solar blackbody emission, a frequency dependent model of the emission must be made so it can be subtracted out with MS-CLEAN or peeling methods.  Non-thermal emission can also occur from transient events such as solar radio bursts} \cite{Reames2013}, \add[AH]{yielding a signal orders of magnitude more intense than that of the quiet Sun or corona.  Transient emission is difficult to characterize to the 1 Jy level, so it is assumed that any data flagged to contain a transient source will be removed for the analysis of the synchrotron emission.}

\subsection{Transients}

In the following subsections, we will review transient emission sources from Earth only, as they will be the most likely to be in the same imaging plane as the synchrotron emission.  

\subsubsection{Auroral Emissions}

Auroral Kilometric Radiation (AKR hereafter) is typically found in frequencies from 50-500 kHz and can sometimes go up to 800 kHz.  AKR is a powerful natural radio source emitting $10^7$ to $10^8$ W, and can exceed $10^9$ in some events \cite{Gurnett74}.  It is typically generated at magnetic latitudes greater than 65\degree at altitudes from 5000-15000 km.  Its power generally increases with magnetospheric activity, especially when substorms develop.  Reported in \cite{Gurnett74}, the Interplanetary Monitoring Platform (IMP) 8 satellite observed AKR at a distance of 25.2 RE on December 20 1973 with a peak emission of $10^{-14}$ W/m$^2$/Hz at 100-200 kHz.  Applying a $r^{-2}$ law, this predicts spectral flux densities of $\sim 1.4\cdot10^{-15}$ W/m$^2$/Hz = $10^{11}$ Jy at the position of the Moon.  While this emission is transient, it far outshines the Radiation Belt emission. 

The source of this emission is thought to be the electron maser instability \cite{Wu79}.  The cyclotron maser mechanism provides the following characteristic predictions: (1) emission occurs near the local electron cyclotron frequency $\Omega_e$ defined in Equation \ref{eqn_cyclo}; (2) the plasma frequency $\omega_{pe} = \sqrt{\frac{n_ee^2}{m_e \epsilon_0}}$ for electron density $n_e$, elemental charge $e$, electron mass $m_e$, and permittivity of free space $\epsilon_0$ in the source region must be much smaller than $\Omega_e$; (3) generation of the radiation occurs primarily in the right-hand extraordinary (R-X) mode.  There is now evidence to back all of these features in the form of an identification of an AKR source region by \citeA{Calvert81}.

\citeA{Mutel08} used data from the 4 spacecraft Cluster array to determine a typical AKR angular beaming pattern.  They found that individual events were highly confined latitudinally (typically $\pm 20\degree$ from the magnetic field tangent direction), but much wider longitudinally, i.e., along the cavity.  The emission is also subjected to strong refraction upwards as it travels, implying that not every event will be detectable from lunar orbit.  By looking at the average beaming of the emission over many days worth of events, we can predict what the emission may look like from the lunar surface.  

\citeA{lamy_GRL_10a} provides a statistical study of AKR as seen from Cassini as it passed by Earth in 1999.  Using data out to several thousand $R_E$, they observe an average beaming of the Northern and Southern AKR consistent with conical beams each tilted towards the nightside, illuminating approximately a hemisphere each, with only sporadic observations from the day side.  Past the shadow zone below 12 $R_E$ on the nightside, emission from both poles is seen at magnetic latitudes lower than 12$\degree$ or so \cite[Fig. 2]{lamy_GRL_10a}.  Since the Moon's orbit is inclined $\sim 28.5\degree$ relative to the Earth's magnetic Equator, this means that observations of AKR from the Moon are predicted to sample all 3 regions: only RH emission from the north pole, only LH emission from the south pole, and a combination of both when its orbit is near the Earth's magnetic equator.   \citeA{lamy_GRL_10a}  reports AKR occurence rates for this region having a nonperiodic average recurrence time of 2-4 hours, with each burst lasting 1-3 hours.  The bursts are distributed log-normal in power with the likeliest power being ~$10^7$ W, which is in agreement with the CMI triggering process.  This implies that AKR can be expected to corrupt measurements of the synchrotron emission about 50\% of the time on the nightside, so roughly 25\% overall.

Related to AKR is auroral hiss, reviewed in \citeA{Sazhin93}.  This is mostly recorded in the evening and night hours in the auroral oval region.  The continuous auroral hiss stays below 30 kHz.  The impulsive auroral hiss is in the 100s of kHz range and can sometimes go up above 500 kHz, usually lasting less than 5 minutes.  The likely main energy source of auroral hiss emissions is electrons at energies below 100 eV at heights greater than about 5000 km above the aurora/ ionosphere precipitating downward.  Maximum spectral flux densities of $\sim$ 1$\cdot10^{-11}$ W/m$^2$/Hz  = $10^{15}$ Jy seen from elevations of 1-2 $R_E$ by satellites such as Injun-5 and Alouette-2 from 2500 km above the surface.  This scales to a maximum spectral flux density of around 6.1$\cdot10^{-18}$ W/m$^2$/Hz = 6.1 $\cdot10^{8}$ Jy at lunar distances.  \citeA{Ondoh91} shows the space based occurrence rates of auroral hiss from the ISIS-2 satellite being between 30-50\% of the time, depending on latitude and geomagnetic local time.

Medium Frequency (MF) bursts are also prominent sources near these frequencies.  MF bursts are correlated with auroral hiss and they are both thought to be associated with the substorm expansion phase \cite{LaBelle97}.  They have a frequency range of about 1.5-4.3 MHz, and usually last around 10 minutes, though they are actually made up of many wave packets lasting 200-300 microseconds each.  Assuming a source altitude of 500 km, on ground brightest packets yield 1-2 microvolt/m/$\sqrt{\text{Hz}}$, but over 100 ms, the average signal is at most 750 nanovolts/m/$\sqrt{\text{Hz}}$.  The wave packet nature of MF Bursts may be due to nonlinear wave processes or bursty characteristics in the precipitating auroral electrons.  The maximum spectral flux density at the lunar surface would be a couple orders of magnitude below that of AKR at around $10^{-18}$ W/m$^2$/Hz  = $\cdot10^{8}$ Jy .  \citeA{LaBelle97} reports the occurrence rates of MF bursts as once every 6 to 20 hours, depending on $K_p$.     

Auroral roar is another class of low frequency emission that is usually found between 2.8 and 3.0 MHz and only has a bandwidth of a few hundred kHz.  It is highly structured and induces voltages of about 1-2$\cdot10^{-13}$ V$^2$/m$^2$/Hz \cite{LaBelle95} and lasts around 10 minutes.  They are thought to occur at about twice the local electron cyclotron frequency, at an altitude of around 250 km.  This emission has a typical strength of 1 microvolt/m, and may be beamed.  AKR in same place is 10-100 millivolts/m implying that auroral roar's total flux density is couple orders of magnitude below that of AKR at around $10^{-18}$ W/m$^2$/Hz  = $\cdot10^{8}$ Jy.  \citeA{Hughes98} reports on the latitudinal dependence for auroral roar occurrence rates, showing that it occurs once every 3-5 hours, and is correlated with $K_p$.

These last 3 sources are sometimes highly localized, with a signal decrease of 35 dB between observations 200 km away \cite{LaBelle97}.  This indicates there may some inherent beaming or directional scattering in these processes that may further decrease the signal seen from the lunar near side.  There also may be a degree of absorption from the ionosphere between the signal source and the lunar surface.  A pathfinder antenna on the lunar near side would be helpful in quantifying how many of these events are detectable from the lunar surface, and how strong they are.

\subsubsection{Terrestrial Continuum Emission}

\citeA{Morgan91} analyzes data from the Dynamics Explorer 1 (DE1) spacecraft and provides an overview of Terrestrial Continuum Emission (TCE).  Virtually all continuum events have their sources near the magnetic equator between 2.0 and 4.0 Re geocentric distance and occur at frequencies between 30 and 200 kHz, with little emission expected at angles less than $20\degree$ from the magnetic equator.  The radiation is beamed outward in a broad beam directed along the magnetic equator with a beam width of about $100\degree$.  

DE1 was 5 Earth Radii away from the Earth, and about 2 Earth Radii away from sources that were more than 2-4 orders of magnitude above the Galactic background.  This implies at lunar distances the flux densities will be 0.1\% of the brightness at DE1, on the order of the Galactic background around $10^{-21}$ W/m$^2$/Hz  = $\cdot10^{5}$ Jy. Unlike the tight beaming of auroral transients in the previous subsection, TCE's wide, equatorial beaming ensures that a lunar near side array would see the majority of TCE events occuring on the visible half of Earth.  \citeA{Morgan91} also reports the occurrence frequency of TCE as 60\% of the time.  The occurrence rates increase sharply at the midnight meridian, and increases toward the dawnward direction.

\subsubsection{Overresolution of Bright Transients}

\add[AH]{For traditional optical telescopes, the resolution for a circular aperture of diameter $D$ meters can be calculated using the Rayleigh Criterion in Equation 3} \cite{Rayleigh1879}. \add[AH]{In this equation, $\lambda$ is the observing wavelength, and $FWHM$ is the full width half maximum of the diffraction pattern from the aperture.  The $FWHM$ is a fundamental limit on the resolution of the telescope, where two point sources closer than this limit are seen as a single point source.  For radio interferometry, the furthest distance between any two receivers in an array determines its resolution, taking the place of $D$ in Equation 3, and $FWHM$ is for the synthesized beam instead of an airy disc for optical telescopes.}  

\begin{align}
FWHM &= 1.22 \frac{\lambda}{D} 
\end{align}

\add[AH]{However, for interferometers there are circumstances when information can be gained about sources smaller than this diffraction limit.}  \citeA{Vidal12} shows how localization better than the beamwidth can be achieved provided there is a strong Signal to Noise Ratio (SNR).  This can be used to estimate the degree to which the array can localize any strong transient emissions.  \add[AH]{In Equation 4, }$\Theta_M$ represents the true minimum size of a source that can still be resolved by the interferometer.  $\beta$ is a constant that depends on the exact configuration of the array, but is usually between 0.5-1.0.  \add[AH]{$L_c$}\remove[AH]{$\lambda_c$} is the value of log-likelihood corresponding to the critical probability of the null hypothesis taking a value of 3.84 for a 2 sigma cutoff, and 8.81 for a 3 sigma cutoff.  The null hypothesis in this case is that the source is a true point soruce, so $\Theta_M$ can also be thought of as the largest source that could be confused with a point source for a given SNR, giving a measure of the true resolution of an array.  This measure is given relative to the Full Width Half Maximum (FWHM) in radians of the synthesized beam of the array, which is the regular method of determining the array's resolution depending on the observing wavelength $\lambda$ and the longest projected distance between receivers $D$. 

\begin{align}
\Theta_M &= \beta \left( \frac{L_c}{2(\textit{SNR})^2} \right)^{\frac{1}{4}} \cdot \textit{FWHM}
\end{align}

For bright transients like strong Auroral Kilometric Radiation, this implies that the array will be able to localize in the plane of sky far better than its beamwidth.  In fact, for all of the following transient signals the ability for a high degree of localization from our array would be interesting science topics in themselves.  The level of overresolution possible for a given SNR transient is listed in Table \ref{table_transients}, alongside other relevant quantities such as occurrence rates and frequency ranges.  Transient emission is difficult to characterize to the 1 Jy level, so it is assumed that any data flagged to contain a transient source will be removed for the analysis of the synchrotron emission.  
\begin{table*}[!t]

\caption{
\textbf{ Characteristics of Earth originating transients as seen from a lunar based radio array} } 

\begin{tabular}{|c|c|c|c|c|}
\hline
    \shortstack{Transient \\ Source} &\shortstack{Frequency \\ Range}& \shortstack{Lunar Flux \\Density 1 MHz} & \shortstack{Occurrence \\Rate} & \shortstack{10 km \\ (Over)resolution} \\
    \hline
    \shortstack{Auroral \\ Kilometric \\ Radiation} & 50 - 800 kHz & $10^{10}$ Jy & 50\% on night side & \shortstack{12-24 arcmin \\ at 500 kHz \\ 10x better} \\
    \hline
    \shortstack{Auroral Hiss} & 100 - 600 kHz & $6 \cdot 10^{8}$ Jy & \shortstack{30-50\% \\ $K_p$ correlation} & \shortstack{18 arcmin at \\ 500 kHz} \\
    \hline
    \shortstack{Medium \\Frequency \\ Bursts} & 1.5-4.3 MHz & $10^{6}$ Jy & \shortstack{10 minutes \\ every 6-20 hours \\ $K_p$ correlation} & \shortstack{42 arcmin \\ at 3 MHz} \\
    \hline
    \shortstack{Auroral Roar} & 2.8-3.0 MHz & $10^{6}$ Jy & \shortstack{10 minutes \\ every 3-5 hours \\ $K_p$ correlation} & \shortstack{42 arcmin \\ at 3 MHz} \\
    \hline
    \shortstack{Terrestrial \\ Continuum \\ Radiation} & 30 - 200 kHz & $10^{5}$ Jy & 60\% & \shortstack{N/A \\ low frequency} \\
    \hline
\end{tabular}

 \label{table_transients}

\end{table*}

\subsection{Unavoidable Noise}

These are noise sources that drive the integration time required for a good detection.  There is no way to subtract it out or get around it.  

\subsubsection{Amplifier Noise}
This is receiver dependent noise that will not be fully understood until actual hardware prototypes are built.  \citeA{Hicks_2012} goes through the process of characterizing the noise and impedance of the amplifier and other electronics of the receiver for the Long Wavelength Array antenna.  Similar techniques would be used to analyze the response of our chosen antenna for a lunar based array.  As a stand in, we choose a level of amplifier noise with equivalent flux density of $10^{-20}$ W/m$^2$/Hz/sr.  This was chosen to roughly match the amplifier noise of other space based antennas such as SunRISE and STEREO/WAVES.

\subsubsection{Quasithermal Noise}

Below 750 kHz plasma thermal noise is a non-negligible factor in solar wind conditions, and dominates the noise levels below 500 kHz.  For a lunar surface with an enhanced electron density from photoionization from Solar photon flux on the dayside, this noise can become the dominant factor.  For electrically short antenna, the formula for the induced voltage by these free electrons is given by \citeA{Vernet89} and  \citeA{Vernet13}, where $n_e$ and $T_e$ are the local electron density (cm$^3$), f is the observing frequency, and L the physical length (m) of one boom (or arm) of the dipole.  We assume for each receiver, each boom is 5 m long.

\begin{align}
V^2_{QTN} = 5\cdot10^{-5} \frac{n_e T_e}{f^3 L}
\label{eqn_qtn}
\end{align}

This is the voltage at the ends of the antenna, so the actual received variations will be multiplied by the gain parameter/wave reflection coefficient from impedance mismatch $\Gamma^2$ which we take as $0.5^2$ in our calculations, matching S/WAVES \cite[Eqn. 7]{Zaslavsky11}.  In Equation \ref{eqn_voltageeqn}, $V_r^2$ is the received spectral voltage power, $V_{noise}^2$ is the amplifier noise, $R_r$ is the radiation resistance of the antenna, $\lambda$ is the observing wavelength, and $B_f$ is the average spectral sky brightness.    

\begin{align}
V^2_r = V_{noise}^2 + \Gamma^2V_{QTN}^2 + 2\Gamma^2R_r\lambda^2B_f
\label{eqn_voltageeqn}
\end{align}

In order to apply this formula to estimate the level of quasithermal noise on the lunar surface, we have to have expected values for the electron density and temperature.  There has never been a radio antenna that could measure the true level of quasithermal noise on the surface of the Moon, so we survey the predictions from theory and remote sensing experiments.  The first experiments that provided an estimate of lunar electron density on the surface were observing radio refractions from the crab nebula \cite{Elsmore57} \cite{Andrew64}.  From these measurements they inferred the presence of a lunar ionosphere above the sunlit lunar surface with peak electron concentrations $n_e \approx$ 500 -– 1000 /cm$^3$.  

A few years later, Soviet spacecraft also did a radio refraction timing experiment \cite{Vasil74} \cite{Vyshlov76} \cite{VySav79}.  Luna 19 and 22 estimated radial density profiles from radio refraction timing data finding the surprising result that the lunar surface may host a stable electron density on the order of 1000/cm$^3$ observed on the sunlit side, including regions near the terminator. 

Lunar Prospector data from 1998$-$1999 used a Electron Reflectometer to measure $T_e$ and $n_e$ at altitude ranges of 30$-$115 km.  The Reflectometer collected data for electrons from 7 eV to 20 keV for 19 months \cite{Chandran13}.  On the day side, $n_e \approx$ 8/cm$^3$ and $T_e \approx$ 12 eV.  At the night side $n_e$ decreases exponentially and $T_e$ reaches to 50 eV. On the lunar night side $n_e$ shows a range of 2--0.002/cm$^3$ and $T_e$ has a range of 15-–50 eV.  Lunar surface potential is found to be highly dependent on electron temperature, which varies with solar input, and may be especially dependent on crustal magnetic fields.

A more recent experiment with LRS (Lunar Radio Science) on Kaguya-SELENE by the Japanese space agency has found evidence of transient enhancements in surface electron density around 250/cm$^3$ but only within a solar zenith angle of 60 degrees.  They used radio occultation experiments with multiple spacecraft to probe the lower lunar atmosphere \cite{Imamura12}.  SELENE did not find a large persistent enhancement like Luna over the whole dayside.  An additional factor that may explain the discrepancy is the amount of ultraviolet radiation at the times of the experiments \cite{Stubbs11}.  The F10.7 index is a measure of the noise level generated by the sun at a wavelength of 10.7 cm at the earth's orbit, and acts as a useful proxy for ultraviolet radiation from the Sun.  The F10.7 index was particularly low at 70 solar flux units (1 sfu = $10^{-22}$ W m$^{−2}$ Hz$^{-1}$) during the SELENE mission at solar minimum.  On the other hand, during the Luna 19 and 22 missions the index was in a range between 75--125 sfu.

There have also been several theory driven approaches to estimating electron conditions at the lunar surface.  \citeA{Colwell07} did a calculation of the photoelectron sheath finding a surface electron density of 60/cm$^3$, using a Maxwellian distribution for the photoelectrons.  This may be outdated by \citeA{Mishra18} and \citeA{Sodha14}, which use a more physically motivated half Fermi Dirac (F-D) distribution for velocities of the photoelectrons.  These analyses find a electron densities on the order of 1000/cm$^3$, and up to 7000/cm$^3$ and higher depending on the solar wind input and photoelectric efficiency of the surface.  Both of these theories predict the reduced photon flux in late afternoon or nighttime will lead to a corresponding decrease in electron density.

\begin{figure*}[!ht]
\centering

\includegraphics[width=1.0\textwidth]{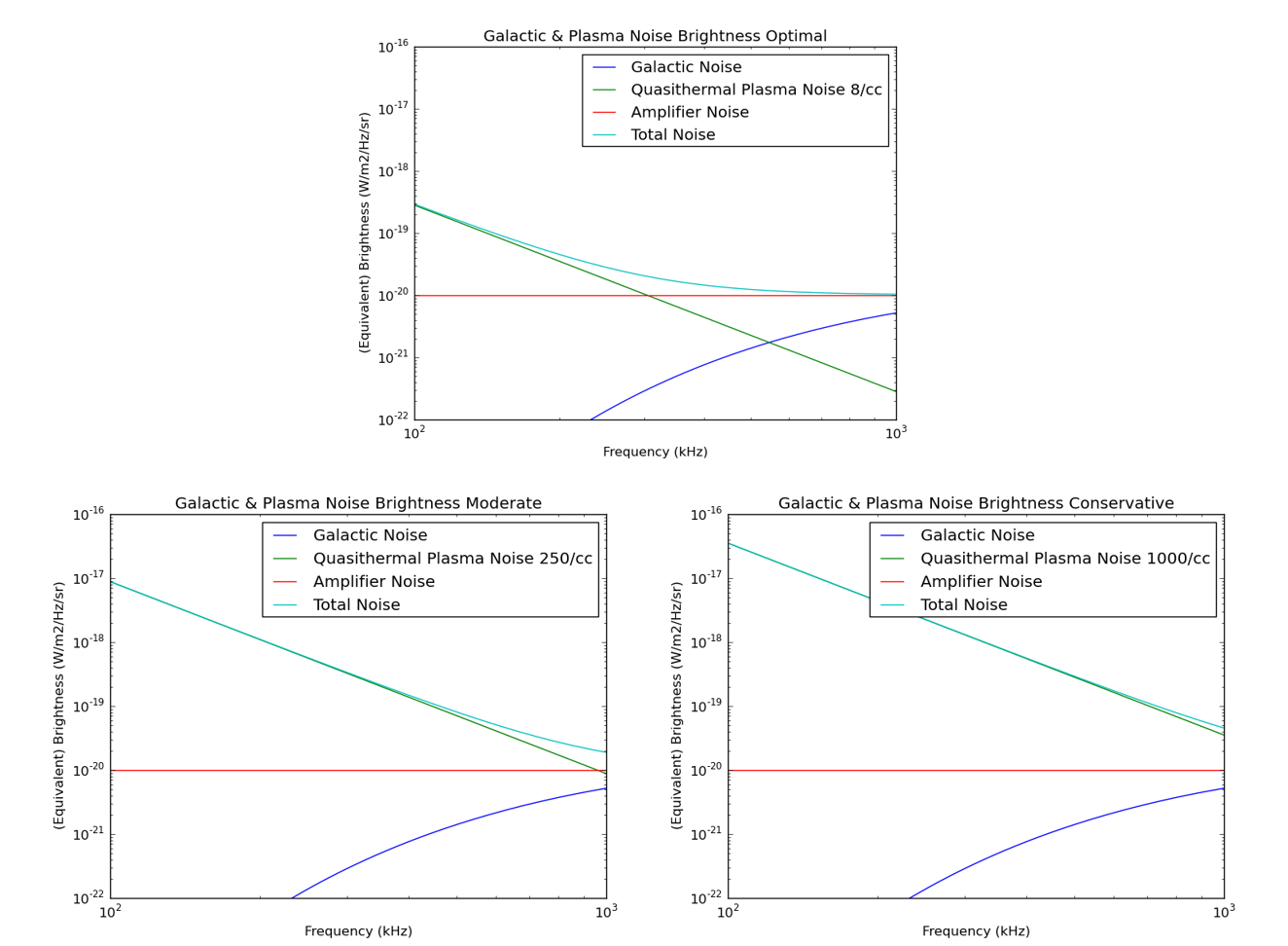}
\caption{\textbf{Noise budgets with different quasithermal noise assumptions.  These include the main unavoidable static noise sources for a lunar surface radio array over the range 100-1000 kHz.  \textit{Top}: Optimal 250/cm$^3$, Amplifier Dominated Noise Budget. \textit{Left}: Moderate, 250/cm$^3$ Electron Quasithermal Noise Dominated.  \textit{Right}: Conservative, 1000/cm$^3$ Electron Quasithermal Noise Dominated.  The sum of these noise sources is multiplied by $4\pi$ steradian to compute the System Equivalent Flux Densities (SEFDs) which we use to compute Signal to Noise ratios.  }}
\label{fig_noiseLevels}
\end{figure*}

We can plug these values into Equation \ref{eqn_qtn} to get conservative (1000/cm$^3$ $n_e$), moderate (250/cm$^3$ $n_e$), and optimistic (8/cm$^3$ $n_e$) values for the plasma noise portion of the noise budget that dominates the lower band.  A electron temperature of 12 eV will be used for all noise budgets, which is justified since the only time it is known to be higher than that is on the night side when $n_e$ is also much lower, so the product of $n_eT_e$ from Equation \ref{eqn_qtn} is equivalent to the optimistic case.  Figure \ref{fig_noiseLevels} shows the equivalent brightness of all the unavoidable noise sources together with a model of the Galactic brightness for reference. 

\subsection{Deciding on an Operational Science Band}

In order to avoid most of the transient sources, we are setting the observing range to 500-1000 kHz.  This range avoids most of the AKR, Auroral Hiss, and Earth Continuum Emission that occurs below 500 kHz, and almost completely avoids the Auroral Roar and MF Bursts that occur above 1.5 MHz.  There were no Salammb\^{o} simulations done to predict the radiation belts above 1.0 MHz, but 1.0-1.5 MHz is likely to be a useful extension of our observing range since there are normally no more transients than there are in the 500-1000 kHz range.  But for the rest of the paper, we assume a operational bandwidth of 500-1000 kHz.  Averaging over this range, the optimistic noise budget gives an average brightness of 1.1$\cdot10^{-20}$ W/m$^2$/Hz/sr which we multiply by $4\pi$ for a system equivalent flux density (SEFD) of 1.38$\cdot10^{-19}$ W/m$^2$/Hz  = $1.38\cdot10^{7}$ Jy.  The moderate noise budget gives an average brightness of 3.66$\cdot10^{-20}$ W/m$^2$/Hz/sr which we multiply by $4\pi$ for a SEFD of 4.6$\cdot10^{-19}$ W/m$^2$/Hz  = 4.6$\cdot10^{7}$ Jy.  The conservative noise budget gives an average brightness of 1.16$\cdot10^{-19}$ W/m$^2$/Hz/sr which we multiply by $4\pi$ for a SEFD of 1.46$\cdot10^{-18}$ W/m$^2$/Hz  = $1.46\cdot10^{8}$ Jy.  If there are any transients that leak into this operating range, we will have to have some system to recognize the extra flux, and filter the data from that bandwidth and time period from the data that will go into the synchrotron imaging.  The data could be processed at high spectral resolution to flag interference before integrating across the observing band for imaging.  

\section{Designing a Mock Array}

Predicted brightness maps have to be run through simulated lunar arrays with realistic noise to see what array size/ configuration will be needed to image the emission of the belts.  However, traditional radio astronomy software is hard coded to assume an Earth based array.  To circumvent this, we manually calculate the antenna separations and insert them along with the simulated visibilities into a Common Astronomy Software Applications (CASA) Measurement Set (MS) file for analysis \cite{Casa07}.  These MS files contain the information of the array configuration, alignment with the sky, and visibility data.  This is a standard format that can be used with a wide range of existing imaging and analysis algorithms.  

The mathematics and theory of creating images with radio arrays has been fleshed out in classic textbooks such as Thompson et al.'s Interferometry and Synthesis in Radio Astronomy \cite{Thompson86}.  Stated informally, the basic insight to understand is that for a group of antennas, the cross correlation of any pair of antennas (a \textit{visibility}) will yield the information of a single 2D Fourier coefficient of the sky brightness pattern.  The exact spatial 2D wave that is sampled depends on the separation between the given pair of radio receivers in units of wavelength of the observing frequency.  The further apart the receivers are in a certain coordinate system oriented towards the imaging target, the higher the spatial frequency sample will be provided, giving higher resolution details at small scales.  Conversely, the closer a pair of receivers are in that same reference frame, the lower the spatial frequency sampled, yielding larger scale structure information at a lower resolution. 

In order to solve for the antenna separations, or \textit{baselines}, a set of locations were chosen using data from the Lunar Reconnaissance Orbiter (LRO) \cite{Chin2007}.  We use  Lunar Orbiter Laser Altimeter (LOLA) data \cite{Barker16} which provides high-resolution Lunar Topography (SLDEM2015) data, giving the altitude for any given longitude and latitude.  The data is in the Moon Mean Earth/Polar Axis (ME) frame, which has the Sub-Earth point at Longitude $0\degree$ Latitude $0\degree$.  The Moon ME frame is standard for all lunar data in the Planetary Data System (PDS).  We use SPICE \cite{SPICE96} to align the Moon ME frame to the celestial sky in order to track its relative position with the Sun and Earth.  By having the array near the sub-Earth point, the array will be very close to planar all the time due to the orbital lock of the Moon with Earth.  The Earth will be directly overhead near the center of the sky at all times, with only slight variations in the projected baselines from the wobbling of the lunar rotation, which is accurately tracked by SPICE.  

With this simulation pipeline in hand, the simulated synchrotron map may be propagated through a model of a distributed radio array on the Moon to produce dirty images that approximate the performance of the array.  A dirty image is an array's imperfect representation of the true sky brightness pattern that has been corrupted by the inherent sparseness of a distributed radio array.  The configuration of the array determines the dirty beam, or the point spread function (psf), that is the combination of unweighted Fourier samples obtained from each pair of antennas.  The dirty image is mathematically equivalent to a convolution of the true sky brightness pattern with the dirty beam.  Any sidelobes or imperfections in the beam will translate into imperfections in the dirty image.

\begin{figure*}[p]
\centering
\includegraphics[width=1.0\textwidth]{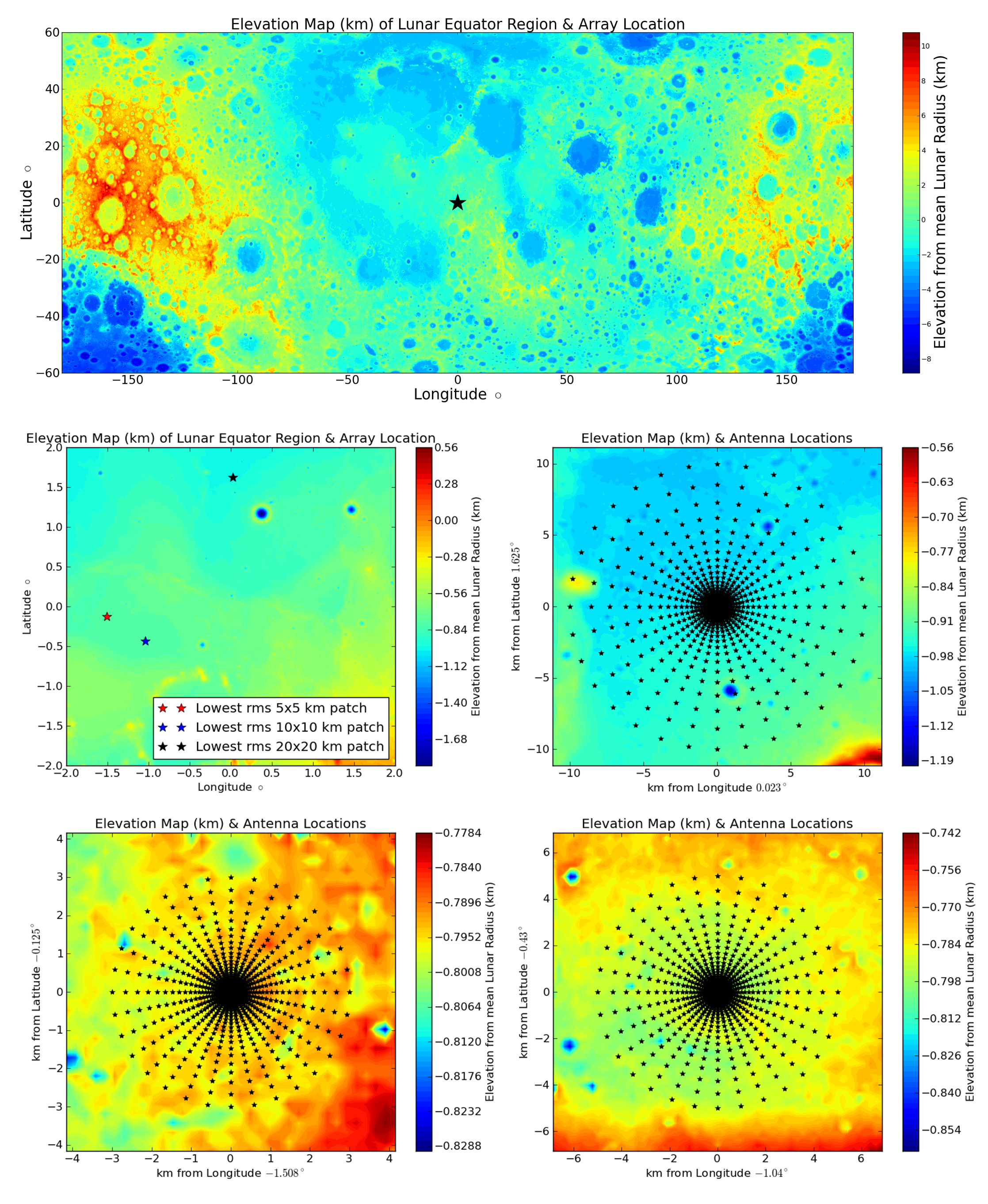}
\caption{\textbf{\textit{Top:} Center of array at sub-Earth point, 0$\degree$ Longitude, $0\degree$ Latitude in the Mean Earth/Polar Axis (ME) frame used for all modern lunar data.  An array near here will have the Earth in the zenith of its sky continuously.  \textit{Middle Left}: Lowest elevation variation array location candidates near the Sub Earth Point for 6, 10, and 20 km arrays.  \textit{Middle Right}:  10 km radius Array, Elevation $\sigma$ = 13.5 m.  \textit{Lower Left}: 5 km radius Array, Elevation $\sigma$ = 5.6 m.  \textit{Lower Right}: 3 km radius Array, Elevation $\sigma$ = 2.8 m.  These elevation maps show different 1024 element array configurations of logarithmically spaced concentric circles.  This configuration is relatively unoptimized, but provides many short baselines where most of the signal for diffuse structures are.  The logarithmic aspect also provides some non-uniformity, increasing the array's \add[AH]{$(u, v)$} \remove[AH]{UV}coverage.}}
\label{fig_arrayMaps}
\end{figure*}

\subsection{Array Locations}

We will test out 3 array sizes: 6 km diameter, 10 km diameter, and 20 km diameter.  In order to find a good place for the center of each array, we zoom in on the area around the Sub Earth point at $0\degree$ Longitude $0\degree$ Latitude.  We limit our search to the area of $\pm 2\degree$ Longitude and Latitude around the Sub-Earth point.  At the equator, each degree of Longitude is 29.67 km, so the approximate area considered was 14085 km$^2$.  Within this area, patches of land with low variance in elevation were found in order to base various sized arrays.  We found the 5$\times$5 km$^2$, the 10$\times$10 km$^2$, and the 20$\times$20 km$^2$ patches that had the lowest variance in elevation according to the SLDEM2015 data with a resolution of 128 pixels per degree.  These locations and their root mean square (RMS) in elevation are shown in Figure \ref{fig_arrayMaps} (b)-(e).

\subsection{Array Formation}

Now that we have locations for the arrays, we have to decide on the configuration of the array.  We assume that we are using 5 m dual-polarization dipole antennas for all our receivers, and that there is a minimum distance of 15 m between receivers.  This limits the maximum density of receivers to $\sim$ 4400 antennas/km$^2$.  Though that dense of a distribution won't be needed everywhere, a large amount of receivers are needed to detect a low frequency synchrotron emission signal that is at least 5 orders of magnitude below the noise.

There exist several algorithms for the optimization of array configuration for a given number of antenna and location.  Iterative algorithms for specific topographies \cite{Boone01} and imaging targets \cite{Boone02} may be used to find a high performing configuration better than simple arrangements such as logarithmically spaced circles.  These techniques may be extended in different ways to take obstacles such as craters into account \cite{Girard13thesis}, or minimize certain parameters like cable length \cite{Zyma17}.  These cables are used to transmit data from each receiver to a central facility for data processing and transmission.  

Minimizing cable length helps decrease construction costs, but an alternative to using cables in the first place is to have a central tower that has a Line of Sight (LOS) view of every antenna that would facilitate communication via a higher frequency antenna.  The equation for the horizon distance is $d = \sqrt{h(2R+h)}$ for radius R and height of observation h.  For lunar radius 1,737.5 km and d = 10 km, this equation can be solved for h = 28.8 meters.  So a tower roughly 30 m or 100 ft tall could be seen by every antenna station out to 10 km.  Though to actually transmit data at an acceptable rate it would need to be taller since transmitting directly to the horizon leaves little room for error.  Fortunately, monopole towers up to 200 feet are commonly used on Earth for a myriad of uses, including wireless communication.  These towers have a small footprint and foundation, and are relatively fast and easy to erect.

The decision for the configuration of a radio array should also take the point spread function into account and assure that an array has sufficient \add[AH]{$(u, v)$} \remove[AH]{UV}coverage.  It has been shown that non-regular arrays such as hierarchical arrays that introduce small tweaks into their array geometry can give better signal to noise ratios or less sidelobe interference than more uniformly spaced arrays\cite{Keto2012}.  Previous experiments with array design have also showed one can employ a sequential optimization strategy to your layout and reach near theoretical limits on sidelobes \cite{WoodyALMA390} \cite{WoodyALMA389}.

Another powerful technique that might be utilized for the configuration of a large scale lunar array is hybrid arrays.  These are getting more popular on the ground with low frequency telescopes like the MWA \cite{MWA13}, LOFAR \cite{LOFAR2013}, and LWA \cite{LWA2009} all employing a version of this strategy.  Hybrid arrays consist of a mixture of single elements and clusters of elements that have been phased up to act as a single element.  Nearby groups of antennas are made to act like a single phased array, and then one employs interferometry to use many of these groups of antennas that spread far away from each other.  This yields both short and long baselines while maintaining a tractable way to handle all the data processing that's spread over many kilometers.    

As an initial stand in for a more optimized array design, we opt for an array shape of logarithmically spaced circles.  By logarithmically spacing the antennas in each arm of the array, more baselines are concentrated in the shorter ranges that provide more signal for imaging the diffuse synchrotron emission belts.  The logarithmic aspect of the layout also adds a layer of non-uniformity to the design, increasing the array's \add[AH]{$(u, v)$} \remove[AH]{UV}coverage.  We simulate a 1024 element array with 32 arms with 32 logarithmically spaced antennas each, and calculate the noiseless visibilities from the synchrotron brightness model.  We did this for a 6 km, 10 km, and 20 km array to see the noiseless response of different synthesized beam responses.  The 1024 element layouts are seen over their respective lunar location in Figure \ref{fig_arrayMaps}.  A more refined optimization of the array configuration that takes into account specific lunar geometries, cable length, point spread functions, and more is left for future work, and is described briefly in the Future Work section of the paper.

\begin{figure*}[!ht]
\centering
\includegraphics[width=1.0\textwidth]{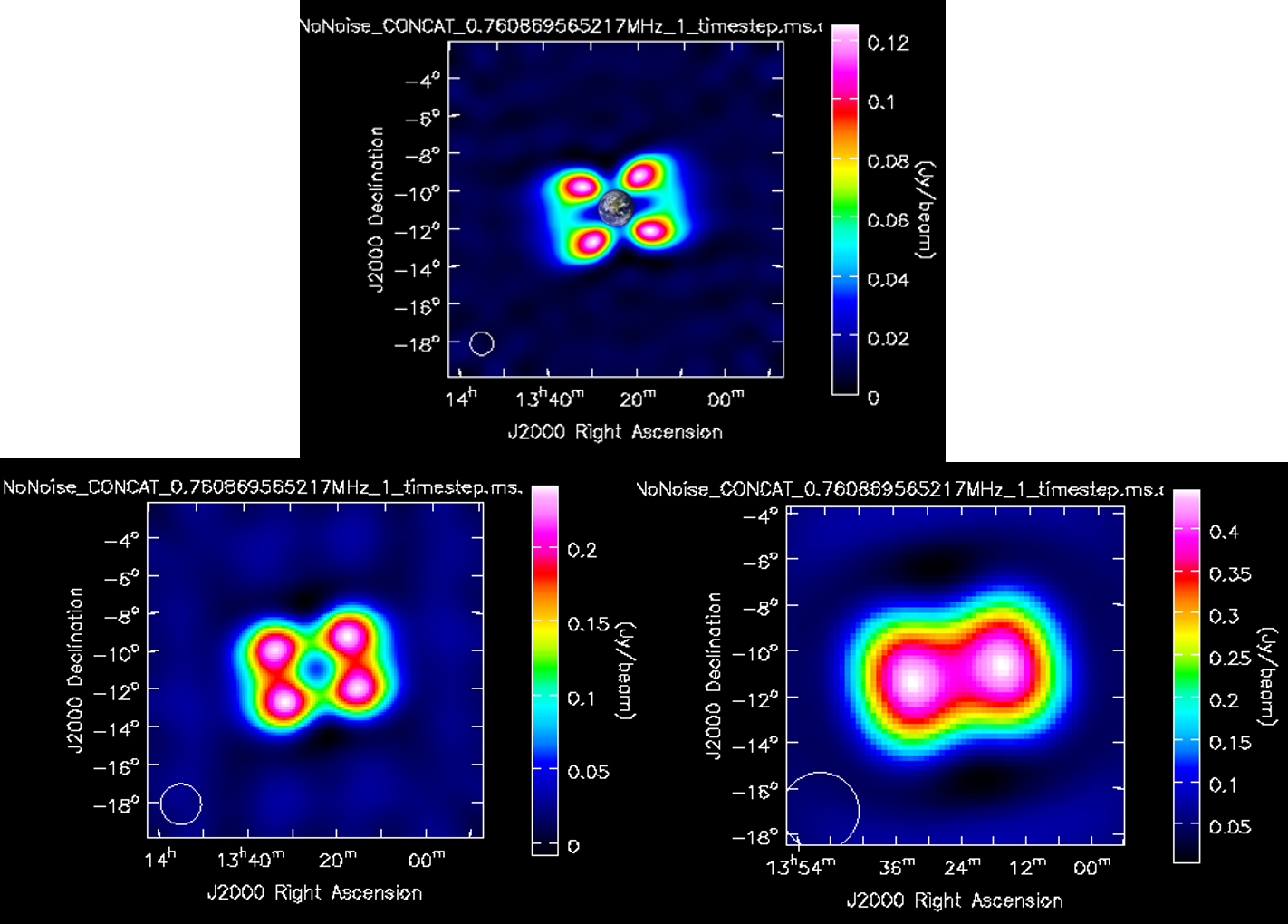}
\caption{\textbf{Noiseless Response of Different Sized Arrays to Synchrotron Emission of Stormy Radiation Belts.  \textit{Top}: Noiseless response of 20 km array.  The 1.91$\degree$ Earth is added in for a scale indicator.  \textit{Left}: Noiseless response of 10 km array.  \textit{Right}: Noiseless response of 6 km array.  Images were made with a Briggs weighting scheme with a robustness parameter of -0.5.}}
\label{fig_noiselessDirty}
\end{figure*}

\subsection{Imaging Performance}

The noiseless recovered images of $\sim$ 2 Jy stormy periods are seen in Figure \ref{fig_noiselessDirty}.  An important thing to note is the maximum of the colorbars in each of the panels.  As the array is made smaller, the beam grows, reducing the resolution of the recovered image, but also making the features brighter because the beam takes in more signal.  The sweet spot may be an array of 10 km since at that resolution 4 main synchrotron lobes are resolved unlike the 6 km array, but the lobes are twice as bright (albeit less well separated) than for the 20 km array.  Images were made with a Briggs weighting scheme with a robustness parameter of -0.5, focusing more on resolution than noise reduction.

Now we add realistic noise to the radio visibilities.  From \cite{Taylor99}, the interferometric noise for a single polarization can be calculated with

\begin{align}
    \sigma = \frac{SEFD}{\eta_s\sqrt{N_{ant}(N_{ant}-1)\Delta \nu \Delta T}}
    \label{eqn_noise}
\end{align}

\begin{figure*}[!hb]
\centering
\includegraphics[width=1.0\textwidth]{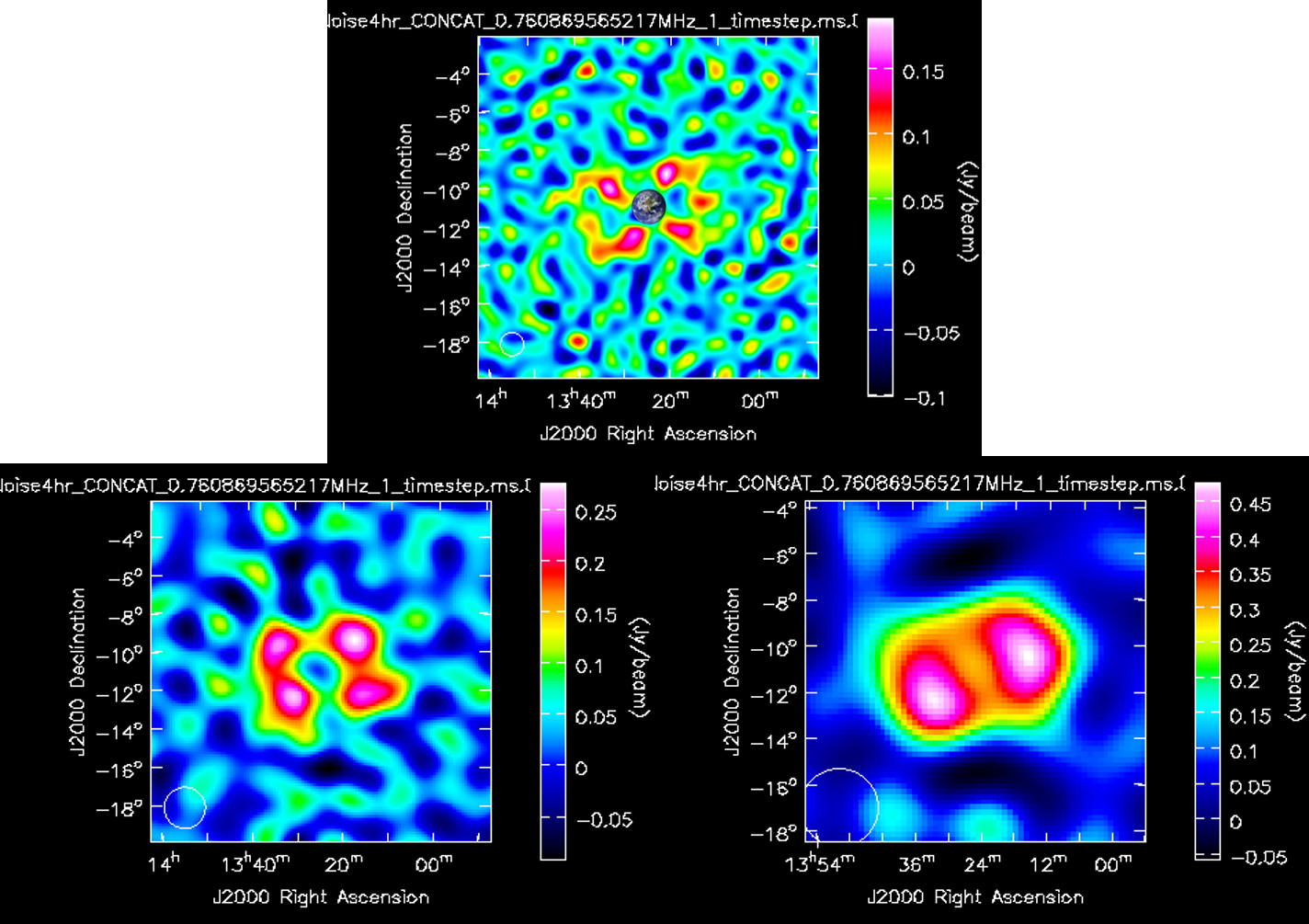}
\caption{\textbf{Recovered Dirty Images after 4 hours Integration with Optimal, Amplifier Limited Noise.  \textit{Top}: Noisy response of 20 km array, $\sigma$ = .0318 Jy/beam $\implies$ SNR $\approx$ 3.93 for each lobe.  The 1.91$\degree$ Earth is added in for a scale indicator.  \textit{Left}: Noisy response of 10 km array, $\sigma$ = 0.041 Jy/beam $\implies$ SNR $\approx$ 5.85 for each lobe.  \textit{Right}: Noisy response of 6 km array, $\sigma$ = 0.073 Jy/beam $\implies$ SNR $\approx$ 6.44 per lobe.  Images were made with a Briggs weighting scheme with a robustness parameter of -0.5, and are showed here completely unCLEANed.  }}
\label{fig_noise}
\end{figure*}

$\eta_s$ is the system efficiency or correlator efficiency, which we have conservatively assumed to be 0.8.  This efficiency is a function of how the correlator does its quantization, with more levels of quantization leading to less signal loss, but more computation with increasing sample rates.  \citeA{Thompson07} provides a table of this correlator efficiency for a number of quantization levels, showing that for a Nyquist sampled voltage waveform, anything over 3 level quantization will lead to a correlator efficiency of over 0.8.  This should not be a limiting factor since modern arrays such as the Very Large Array (VLA) use 8 bit sampling, leading to 256 quantization levels, and a correlator efficiency over 0.9.  The System Equivalent Flux Density (SEFD) is a useful way to talk about a radio antenna's total noise because it ties in both the effective area and the system temperature, giving a simple way to compare the signal and the noise.  We take the SEFD as the average noise over our operational science band described in section 3.4 and take $\Delta \nu$ to be 500 kHz.

Equation \ref{eqn_noise} is for point source sensitivity, and is valid because the Fourier transform of a delta function has a constant non-zero amplitude.  For diffuse sources in the sky such as the radiation belts, the distribution of baselines is important.  For the synchrotron emission belts, Figure \ref{fig_truth} (b) shows that most of the power is in a couple 10s of wavelengths, with generally more power the shorter the baseline.  This means the amount of signal added from a baseline is not constant, and imaging software like CASA is needed to understand what the SNR would be for a given array configuration imaging diffuse structures such as the synchrotron emission from radiation belts.  The units for the Signal and Noise in the recovered images from an interferometer are Jy/beam.  Figure \ref{fig_truth} (c) also shows that the 3 areas of \add[AH]{$(u, v)$} \remove[AH]{UV}space that the most power have phases close to either 0$\degree$, $180\degree$, or $-180\degree$.  This will be a useful check on the real measurements, and may be used to detect errors in phase measurement.  However, we will not go into this advanced level of array calibration for this paper.

We treat the values with an appropriate amount of noise, using Equation \ref{eqn_noise} with $N_{Ant}$ = 2 for each visibility.  Equation \ref{eqn_noise} also tells us that overall noise decreases roughly linearly with increasing $N_{Ant}$.  We can use our 1024 element model to estimate a 16384 element array by dividing the noise by 16, as long as we assume the expanded array has a similar distribution of baselines.  Under this assumption, a 16384 element array has $\sim$256 similar baselines for every 1 baseline of a 1024 element array.  So when adding all the visibility data to create an image, the Fourier sample for that baseline will have its noise decreased by a factor of $\sqrt{256}=16$ when compared to the single corresponding baseline for a 1024 element array.  So by dividing the noise from our 1024 element arrays by 16, we have simulated a 16384 element array spread over 6, 10 and 20 km.  Recovered dirty images of stormy period synchrotron emission for a 4 hour integration time using 16384 receivers in an optimal, amplifier limited noise environment are seen in Figure \ref{fig_noise}.  With the SNR in Jansky/beam from these simulations, we can come up with predicted times it would take to reach a given SNR for a particular noise environment.  Data were imaged using a Briggs weighting scheme \cite{Briggs99} with a robustness parameter of -0.5, so more on the uniform weighting side as opposed to natural weighting.  This seemed to make the best images for this imaging target, with larger robustness values giving up too much resolution, while more negative values being noisier.  The images \add[AH]{have not been deconvolved in any way.} \remove[AH] {are otherwise uncleaned, with 0 iterations of any sort of CLEANing algorithm.}

From Figure \ref{fig_brightness}, the average integrated spectral flux density for the radiation belts in a noisy storm period is 2 Jy over 500-1000 kHz, while for a calm period it is 1.4 Jy.  The translates into needing an integration time roughly twice as long in order to reach a similar SNR for a given array.  There is roughly a factor of 3.3 between optimal and moderate noise, and a factor of 3.16 between moderate and conservative noise.  This means that to reach the same SNR takes $\sim$10 times longer in moderate environment than in the optimal, and also (at least) 10 times longer in the conservative regime over the moderate environment.  So if the antenna array can be powered during lunar night, snapshots could be taken of the radiation belts every couple hours.  On the other hand, electron densities over 1000/cm$^3$ at low Solar zenith angles (near lunar noon) could overwhelm the array to the small signal that the radiation belts give off.  The expected integration times for a 16384 element array for our various noise budgets and array sizes is shown in Table \ref{table_integration}.

\begin{table*}[!ht]

\caption{
\textbf{ Expected Integration Times for 16384 Element Arrays of Various Sizes} } 

\begin{tabular}{|c|c|c|c|}
\hline
     \shortstack{Integration Time (minutes) for\\ 16384 Element Array over 500 kHz}& 6 km array & 10 km array & 20 km array  \\
     \hline
    Optimal Noise 3 $\sigma$ Lobe Detection Calm &  104 & 126 & 280 \\
    \hline
    Optimal Noise 3 $\sigma$ Lobe Detection Storm & 52  & 63 & 140 \\
    \hline
    Moderate Noise 3 $\sigma$ Lobe Detection Calm & 1132  & 1372  & 3050 \\
    \hline
    Moderate Noise 3 $\sigma$ Lobe Detection Storm & 566  & 686 & 1525 \\
    \hline
    Conservative Noise 3 $\sigma$ Lobe Detection Calm &  11096 & 13442 & 28873  \\
    \hline
    Conservative Noise 3 $\sigma$ Lobe Detection Storm &  5548 & 6721  & 14936  \\
    \hline
\end{tabular}

 \label{table_integration}

\end{table*}

\section{Discussion}

Imaging the synchrotron emission from the Earth's radiation belts at regular intervals would go a long way towards understanding the global response of Earth to variable Solar input.  However, due to the relative weakness of the signal compared to the unavoidable noise sources from the lunar ionosphere and receiver electronics, thousands of antennas would be needed to get good measurements at a decent cadence.  This paper outlines many of the transient noise sources and provides estimates of what it would take to achieve useful results, but it is only a first attempt at answering the problem.  Many antenna design \& implementation details would have to be taken into account for a real mission, a few of which are listed in the Future Work section.

Table \ref{table_integration} outlines the integration times needed for successful detections of the synchrotron emission under different conditions, saying that the data equivalent needed for a certain level of detection is X minutes times 500 kHz.  There is an implicit optimism here because in reality, a flagging system would need to be implemented that could take out noisy channels that have other sources of unknown strength overpowering the synchrotron signal.  This would mean it would likely take longer than stated in the table to actually reach the amount of data needed for a given SNR.  Another important factor not mentioned so far is duty cycle.  Most antennas are not recording data 100\% of the time.  A system where every antenna has a 50\% duty cycle means that it would take twice as long to collect the same amount of signal. 

Some useful takeaways from this paper are that $\sim$10 km seems to be a good compromise in array size because at that resolution 4 main synchrotron lobes are resolved, but are still relatively bright.  A 10x10 km patch could hold over 440000 antennas if densely packed, but a circular distribution with many logarithmically spaced arms could make do with 16384 elements.  Several low variance altitude regions near the Sub Earth point of the lunar surface were identified as promising array locations.  This work also demonstrates a data processing pipeline combining SPICE, lunar surface data from LRO, and CASA that can generate the dirty images for a lunar array.  The integration times required for detections are predicted to be highly dependent on Solar Zenith Angle, since less incident Sunlight will lead to fewer photoionized electrons, which will mean less quasithermal noise.  This results in faster snapshots of the synchrotron emission as you move from lunar noon, to lunar late afternoon, to lunar night.  This provides an incentive for a power supply system that can power the system as late as possible into the lunar night.  This will require either highly efficient solar panels and batteries, or a radioisotope thermoelectric generator.

\section{Future Work}

As discussed in the Array Formation section, there are a number of optimizations that could be made to the array configuration.  Logarithmically spaced circles are used as a stand-in, but in reality we would want to optimize the configuration, avoiding any small craters at the array site, minimizing key parameters such as total cable length, and designing the point spread function to have good \add[AH]{$(u, v)$} \remove[AH]{UV}coverage.  In addition to increasing the imaging performance of the array, these optimizations can also help decrease construction costs.  

Improvements in the simulations could be made by including a channel dependent simulated foreground removal for removable constant noise sources such as blackbody signals and Galactic background structure.  The data processing pipeline could also use a fleshed out transient event detection scheme that removes flagged channels from the data that goes into the synchrotron emission imaging.  The pipeline could then be tested on imaging these transient signals to demonstrate the degree of localization possible for a given SNR transient. 

As discussed in the Amplifier Noise section, there is hardware specific characterization of the noise and impedance of the receiver to be done.  \citeA{Hicks_2012} provides a useful guide to look to as they go through these processes for the Long Wavelength Array (LWA) antenna.  Similar techniques would be used to analyze the response of our chosen antenna for a lunar based array.  Mutual coupling and Galactic noise correlation can lead to a decrease in sensitivity for arrays with receivers less than a few wavelengths away from one another, as discussed in \citeA{Ellingson11}.  This is reflected in a increase in expected SEFD for the array, especially for beams formed over $10\degree$ from zenith.  For beams near zenith the effects of coupling is frequency dependent, and may be better or worse than expected. For the purpose of imaging the Earth's synchrotron emission, the consequences from coupling are minimal since the array's location ensures that the Earth will always be near the sky's zenith.  In order to unlock the array's full potential, studies of the expected SEFD as a function of elevation angle and frequency will have to be done, as \citeA{Ellingson11} did for the LWA.     

The NASA SMD recently chose the Radio wave Observations on the Lunar Surface of the photoElectron Sheath (ROLSES) mission with PI Robert MacDowall to put a STEREO WAVES inspired radio antenna on the lunar near side \cite{NASACLPS}.  This will be an excellent pathfinder for many engineering aspects of the array not described in this paper, and will also finally provide direct measurements of the photoelectron sheath density near the surface over the course of the lunar day.  This will solidify the noise budget in Figure \ref{fig_noiseLevels}, and will help drive requirements for signal to noise levels for all future lunar radio arrays.  It will also provide occurrence rates and flux density levels for transient events detectable on the lunar near side.  The instrument will be flown as part of the Commercial Lunar Payload Services (CLPS) program, where private landers will robotically deliver and deploy selected payloads.  The expected Payload Delivery Date is August 2020.  Another CLPS mission is Solar Cell Demonstration Platform for Enabling Long-Term Lunar Surface Power will demonstrate advanced solar arrays for longer mission duration.  The expected Payload Delivery Date is March 2020.

\acknowledgments
Thank you to the Lunar Reconnaissance Orbiter (LRO) and Lunar Orbiter Laser Altimeter (LOLA) teams for mapping the Moon to an unprecedented degree.  This work was directly supported by the NASA Solar System Exploration Research Virtual Institute cooperative agreement number 80ARC017M0006, as part of the Network for Exploration and Space Science (NESS) team.  This work adheres to the common Enabling FAIR data Project guidelines, and offers all data and code up at \url{https://github.com/alexhege/LunarSynchrotronArray}.  We would also like to thank the reviewers for their helpful suggestions.


%
%

\bibliography{agusample.bib}

\begin{thebibliography}{}

\bibitem [\protect \citeauthoryear {%
{Acton}%
}{%
{Acton}%
}{%
{\protect \APACyear {1996}}%
}]{%
SPICE96}
\APACinsertmetastar {%
SPICE96}%
\begin{APACrefauthors}%
{Acton}, C\BPBI H.%
\end{APACrefauthors}%
\unskip\
\newblock
\APACrefYearMonthDay{1996}{{\APACmonth{01}}}{}.
\newblock
{\BBOQ}\APACrefatitle {{Ancillary data services of NASA's Navigation and
  Ancillary Information Facility}} {{Ancillary data services of NASA's
  Navigation and Ancillary Information Facility}}.{\BBCQ}
\newblock
\APACjournalVolNumPages{Planetary and Space Science}{44}{}{65-70}.
\newblock
\begin{APACrefDOI} \doi{10.1016/0032-0633(95)00107-7} \end{APACrefDOI}
\PrintBackRefs{\CurrentBib}

\bibitem [\protect \citeauthoryear {%
Andrew%
, Branson%
\BCBL {}\ \BBA {} Wills%
}{%
Andrew%
\ \protect \BOthers {.}}{%
{\protect \APACyear {1964}}%
}]{%
Andrew64}
\APACinsertmetastar {%
Andrew64}%
\begin{APACrefauthors}%
Andrew, B\BPBI H.%
, Branson, N\BPBI J\BPBI B\BPBI A.%
\BCBL {}\ \BBA {} Wills, D.%
\end{APACrefauthors}%
\unskip\
\newblock
\APACrefYearMonthDay{1964}{}{}.
\newblock
{\BBOQ}\APACrefatitle {Radio observation of the Crab nebula during a lunar
  occultation} {Radio observation of the crab nebula during a lunar
  occultation}.{\BBCQ}
\newblock
\APACjournalVolNumPages{Nature}{203}{}{171-173}.
\newblock
\begin{APACrefURL} \url{https://doi.org/10.1038/203171b0} \end{APACrefURL}
\newblock
\begin{APACrefDOI} \doi{10.1038/203171b0} \end{APACrefDOI}
\PrintBackRefs{\CurrentBib}

\bibitem [\protect \citeauthoryear {%
{Angelopoulos}%
}{%
{Angelopoulos}%
}{%
{\protect \APACyear {2008}}%
}]{%
themis08}
\APACinsertmetastar {%
themis08}%
\begin{APACrefauthors}%
{Angelopoulos}, V.%
\end{APACrefauthors}%
\unskip\
\newblock
\APACrefYearMonthDay{2008}{{\APACmonth{12}}}{}.
\newblock
{\BBOQ}\APACrefatitle {{The THEMIS Mission}} {{The THEMIS Mission}}.{\BBCQ}
\newblock
\APACjournalVolNumPages{Space Science Reviews}{141}{}{5-34}.
\newblock
\begin{APACrefDOI} \doi{10.1007/s11214-008-9336-1} \end{APACrefDOI}
\PrintBackRefs{\CurrentBib}

\bibitem [\protect \citeauthoryear {%
Baker%
\ \protect \BOthers {.}}{%
Baker%
\ \protect \BOthers {.}}{%
{\protect \APACyear {2019}}%
}]{%
Baker19}
\APACinsertmetastar {%
Baker19}%
\begin{APACrefauthors}%
Baker, D\BPBI N.%
, Hoxie, V.%
, Zhao, H.%
, Jaynes, A\BPBI N.%
, Kanekal, S.%
, Li, X.%
\BCBL {}\ \BBA {} Elkington, S.%
\end{APACrefauthors}%
\unskip\
\newblock
\APACrefYearMonthDay{2019}{Mar}{13}.
\newblock
{\BBOQ}\APACrefatitle {Multiyear Measurements of Radiation Belt Electrons:
  Acceleration, Transport, and Loss} {Multiyear measurements of radiation belt
  electrons: Acceleration, transport, and loss}.{\BBCQ}
\newblock
\APACjournalVolNumPages{Journal of Geophysical Research: Space
  Physics}{124}{}{}.
\newblock
\begin{APACrefURL}
  \url{https://agupubs.onlinelibrary.wiley.com/doi/abs/10.1029/2018JA026259}
  \end{APACrefURL}
\newblock
\begin{APACrefDOI} \doi{10.1029/2018JA026259} \end{APACrefDOI}
\PrintBackRefs{\CurrentBib}

\bibitem [\protect \citeauthoryear {%
Baker%
\ \protect \BOthers {.}}{%
Baker%
\ \protect \BOthers {.}}{%
{\protect \APACyear {2014}}%
}]{%
Baker2014}
\APACinsertmetastar {%
Baker2014}%
\begin{APACrefauthors}%
Baker, D\BPBI N.%
, Jaynes, A\BPBI N.%
, Hoxie, V\BPBI C.%
, Thorne, R\BPBI M.%
, Foster, J\BPBI C.%
, Li, X.%
\BDBL {}Lanzerotti, L\BPBI J.%
\end{APACrefauthors}%
\unskip\
\newblock
\APACrefYearMonthDay{2014}{Nov}{26}.
\newblock
{\BBOQ}\APACrefatitle {An impenetrable barrier to ultrarelativistic electrons
  in the Van Allen radiation belts} {An impenetrable barrier to
  ultrarelativistic electrons in the van allen radiation belts}.{\BBCQ}
\newblock
\APACjournalVolNumPages{Nature}{515}{}{531}.
\newblock
\begin{APACrefURL} \url{https://doi.org/10.1038/nature13956} \end{APACrefURL}
\PrintBackRefs{\CurrentBib}

\bibitem [\protect \citeauthoryear {%
Bale%
\ \protect \BOthers {.}}{%
Bale%
\ \protect \BOthers {.}}{%
{\protect \APACyear {2008}}%
}]{%
Bale2008}
\APACinsertmetastar {%
Bale2008}%
\begin{APACrefauthors}%
Bale, S\BPBI D.%
, Ullrich, R.%
, Goetz, K.%
, Alster, N.%
, Cecconi, B.%
, Dekkali, M.%
\BDBL {}Pulupa, M.%
\end{APACrefauthors}%
\unskip\
\newblock
\APACrefYearMonthDay{2008}{Apr}{01}.
\newblock
{\BBOQ}\APACrefatitle {The Electric Antennas for the STEREO/WAVES Experiment}
  {The electric antennas for the stereo/waves experiment}.{\BBCQ}
\newblock
\APACjournalVolNumPages{Space Science Reviews}{136}{1}{529--547}.
\newblock
\begin{APACrefURL} \url{https://doi.org/10.1007/s11214-007-9251-x}
  \end{APACrefURL}
\newblock
\begin{APACrefDOI} \doi{10.1007/s11214-007-9251-x} \end{APACrefDOI}
\PrintBackRefs{\CurrentBib}

\bibitem [\protect \citeauthoryear {%
Barker%
\ \protect \BOthers {.}}{%
Barker%
\ \protect \BOthers {.}}{%
{\protect \APACyear {2016}}%
}]{%
Barker16}
\APACinsertmetastar {%
Barker16}%
\begin{APACrefauthors}%
Barker, M.%
, Mazarico, E.%
, Neumann, G.%
, Zuber, M.%
, Haruyama, J.%
\BCBL {}\ \BBA {} Smith, D.%
\end{APACrefauthors}%
\unskip\
\newblock
\APACrefYearMonthDay{2016}{}{}.
\newblock
{\BBOQ}\APACrefatitle {A new lunar digital elevation model from the Lunar
  Orbiter Laser Altimeter and SELENE Terrain Camera} {A new lunar digital
  elevation model from the lunar orbiter laser altimeter and selene terrain
  camera}.{\BBCQ}
\newblock
\APACjournalVolNumPages{Icarus}{273}{}{346 - 355}.
\newblock
\begin{APACrefURL}
  \url{http://www.sciencedirect.com/science/article/pii/S0019103515003450}
  \end{APACrefURL}
\newblock
\begin{APACrefDOI} \doi{https://doi.org/10.1016/j.icarus.2015.07.039}
  \end{APACrefDOI}
\PrintBackRefs{\CurrentBib}

\bibitem [\protect \citeauthoryear {%
Beutier%
\ \BBA {} Boscher%
}{%
Beutier%
\ \BBA {} Boscher%
}{%
{\protect \APACyear {1995}}%
}]{%
Salammbo95}
\APACinsertmetastar {%
Salammbo95}%
\begin{APACrefauthors}%
Beutier, T.%
\BCBT {}\ \BBA {} Boscher, D.%
\end{APACrefauthors}%
\unskip\
\newblock
\APACrefYearMonthDay{1995}{}{}.
\newblock
{\BBOQ}\APACrefatitle {A three-dimensional analysis of the electron radiation
  belt by the Salammb\^{o} code} {A three-dimensional analysis of the electron
  radiation belt by the salammb\^{o} code}.{\BBCQ}
\newblock
\APACjournalVolNumPages{Journal of Geophysical Research: Space
  Physics}{100}{A8}{14853-14861}.
\newblock
\begin{APACrefURL}
  \url{https://agupubs.onlinelibrary.wiley.com/doi/abs/10.1029/94JA03066}
  \end{APACrefURL}
\newblock
\begin{APACrefDOI} \doi{10.1029/94JA03066} \end{APACrefDOI}
\PrintBackRefs{\CurrentBib}

\bibitem [\protect \citeauthoryear {%
Bolton%
\ \protect \BOthers {.}}{%
Bolton%
\ \protect \BOthers {.}}{%
{\protect \APACyear {2002}}%
}]{%
Bolton2002}
\APACinsertmetastar {%
Bolton2002}%
\begin{APACrefauthors}%
Bolton, S\BPBI J.%
, Janssen, M.%
, Thorne, R.%
, Levin, S.%
, Klein, M.%
, Gulkis, S.%
\BDBL {}West, R.%
\end{APACrefauthors}%
\unskip\
\newblock
\APACrefYearMonthDay{2002}{Feb}{28}.
\newblock
{\BBOQ}\APACrefatitle {Ultra-relativistic electrons in Jupiter\&\#39;s
  radiation belts} {Ultra-relativistic electrons in jupiter\&\#39;s radiation
  belts}.{\BBCQ}
\newblock
\APACjournalVolNumPages{Nature}{415}{}{987}.
\newblock
\begin{APACrefURL} \url{https://doi.org/10.1038/415987a} \end{APACrefURL}
\PrintBackRefs{\CurrentBib}

\bibitem [\protect \citeauthoryear {%
{Boone, F.}%
}{%
{Boone, F.}%
}{%
{\protect \APACyear {2001}}%
}]{%
Boone01}
\APACinsertmetastar {%
Boone01}%
\begin{APACrefauthors}%
{Boone, F.}%
\end{APACrefauthors}%
\unskip\
\newblock
\APACrefYearMonthDay{2001}{}{}.
\newblock
{\BBOQ}\APACrefatitle {Interferometric array design: Optimizing the locations
  of the antenna pads} {Interferometric array design: Optimizing the locations
  of the antenna pads}.{\BBCQ}
\newblock
\APACjournalVolNumPages{A\&A}{377}{1}{368-376}.
\newblock
\begin{APACrefURL} \url{https://doi.org/10.1051/0004-6361:20011105}
  \end{APACrefURL}
\newblock
\begin{APACrefDOI} \doi{10.1051/0004-6361:20011105} \end{APACrefDOI}
\PrintBackRefs{\CurrentBib}

\bibitem [\protect \citeauthoryear {%
{Boone, F.}%
}{%
{Boone, F.}%
}{%
{\protect \APACyear {2002}}%
}]{%
Boone02}
\APACinsertmetastar {%
Boone02}%
\begin{APACrefauthors}%
{Boone, F.}%
\end{APACrefauthors}%
\unskip\
\newblock
\APACrefYearMonthDay{2002}{}{}.
\newblock
{\BBOQ}\APACrefatitle {Interferometric array design: Distributions of Fourier
  samples for imaging} {Interferometric array design: Distributions of fourier
  samples for imaging}.{\BBCQ}
\newblock
\APACjournalVolNumPages{A\&A}{386}{3}{1160-1171}.
\newblock
\begin{APACrefURL} \url{https://doi.org/10.1051/0004-6361:20020297}
  \end{APACrefURL}
\newblock
\begin{APACrefDOI} \doi{10.1051/0004-6361:20020297} \end{APACrefDOI}
\PrintBackRefs{\CurrentBib}

\bibitem [\protect \citeauthoryear {%
Boscher%
, Bourdarie%
, Thorne%
\BCBL {}\ \BBA {} Abel%
}{%
Boscher%
\ \protect \BOthers {.}}{%
{\protect \APACyear {2000}}%
}]{%
Salammbo2000}
\APACinsertmetastar {%
Salammbo2000}%
\begin{APACrefauthors}%
Boscher, D.%
, Bourdarie, S.%
, Thorne, R.%
\BCBL {}\ \BBA {} Abel, B.%
\end{APACrefauthors}%
\unskip\
\newblock
\APACrefYearMonthDay{2000}{}{}.
\newblock
{\BBOQ}\APACrefatitle {Influence of the wave characteristics on the electron
  radiation belt distribution} {Influence of the wave characteristics on the
  electron radiation belt distribution}.{\BBCQ}
\newblock
\APACjournalVolNumPages{Advances in Space Research}{26}{1}{163 - 166}.
\newblock
\begin{APACrefURL}
  \url{http://www.sciencedirect.com/science/article/pii/S0273117799010431}
  \end{APACrefURL}
\newblock
\APACrefnote{Space Weather: Physics and Applications}
\newblock
\begin{APACrefDOI} \doi{https://doi.org/10.1016/S0273-1177(99)01043-1}
  \end{APACrefDOI}
\PrintBackRefs{\CurrentBib}

\bibitem [\protect \citeauthoryear {%
Bougeret%
\ \protect \BOthers {.}}{%
Bougeret%
\ \protect \BOthers {.}}{%
{\protect \APACyear {2008}}%
}]{%
Bougeret2008}
\APACinsertmetastar {%
Bougeret2008}%
\begin{APACrefauthors}%
Bougeret, J\BPBI L.%
, Goetz, K.%
, Kaiser, M\BPBI L.%
, Bale, S\BPBI D.%
, Kellogg, P\BPBI J.%
, Maksimovic, M.%
\BDBL {}Zouganelis, I.%
\end{APACrefauthors}%
\unskip\
\newblock
\APACrefYearMonthDay{2008}{Apr}{01}.
\newblock
{\BBOQ}\APACrefatitle {S/WAVES: The Radio and Plasma Wave Investigation
  on the STEREO Mission} {S/waves: The radio and plasma wave investigation
  on the stereo mission}.{\BBCQ}
\newblock
\APACjournalVolNumPages{Space Science Reviews}{136}{1}{487--528}.
\newblock
\begin{APACrefURL} \url{https://doi.org/10.1007/s11214-007-9298-8}
  \end{APACrefURL}
\newblock
\begin{APACrefDOI} \doi{10.1007/s11214-007-9298-8} \end{APACrefDOI}
\PrintBackRefs{\CurrentBib}

\bibitem [\protect \citeauthoryear {%
Bourdarie%
, Boscher%
, Beutier%
, Sauvaud%
\BCBL {}\ \BBA {} Blanc%
}{%
Bourdarie%
\ \protect \BOthers {.}}{%
{\protect \APACyear {1996}}%
}]{%
Salammbo96}
\APACinsertmetastar {%
Salammbo96}%
\begin{APACrefauthors}%
Bourdarie, S.%
, Boscher, D.%
, Beutier, T.%
, Sauvaud, J\BHBI A.%
\BCBL {}\ \BBA {} Blanc, M.%
\end{APACrefauthors}%
\unskip\
\newblock
\APACrefYearMonthDay{1996}{}{}.
\newblock
{\BBOQ}\APACrefatitle {Magnetic storm modeling in the Earth's electron belt by
  the Salammb\^{o} code} {Magnetic storm modeling in the earth's electron belt
  by the salammb\^{o} code}.{\BBCQ}
\newblock
\APACjournalVolNumPages{Journal of Geophysical Research: Space
  Physics}{101}{A12}{27171-27176}.
\newblock
\begin{APACrefURL}
  \url{https://agupubs.onlinelibrary.wiley.com/doi/abs/10.1029/96JA02284}
  \end{APACrefURL}
\newblock
\begin{APACrefDOI} \doi{10.1029/96JA02284} \end{APACrefDOI}
\PrintBackRefs{\CurrentBib}

\bibitem [\protect \citeauthoryear {%
{Briggs}%
, {Schwab}%
\BCBL {}\ \BBA {} {Sramek}%
}{%
{Briggs}%
\ \protect \BOthers {.}}{%
{\protect \APACyear {1999}}%
}]{%
Briggs99}
\APACinsertmetastar {%
Briggs99}%
\begin{APACrefauthors}%
{Briggs}, D\BPBI S.%
, {Schwab}, F\BPBI R.%
\BCBL {}\ \BBA {} {Sramek}, R\BPBI A.%
\end{APACrefauthors}%
\unskip\
\newblock
\APACrefYearMonthDay{1999}{Jan}{}.
\newblock
{\BBOQ}\APACrefatitle {{Imaging}} {{Imaging}}.{\BBCQ}
\newblock
\BIn{} G\BPBI B.~{Taylor}, C\BPBI L.~{Carilli}\BCBL {}\ \BBA {} R\BPBI
  A.~{Perley}\ (\BEDS), \APACrefbtitle {Synthesis Imaging in Radio Astronomy
  II} {Synthesis imaging in radio astronomy ii}\ (\BVOL~180, \BPG~127).
\PrintBackRefs{\CurrentBib}

\bibitem [\protect \citeauthoryear {%
Calvert%
}{%
Calvert%
}{%
{\protect \APACyear {1981}}%
}]{%
Calvert81}
\APACinsertmetastar {%
Calvert81}%
\begin{APACrefauthors}%
Calvert, W.%
\end{APACrefauthors}%
\unskip\
\newblock
\APACrefYearMonthDay{1981}{}{}.
\newblock
{\BBOQ}\APACrefatitle {The auroral plasma cavity} {The auroral plasma
  cavity}.{\BBCQ}
\newblock
\APACjournalVolNumPages{Geophysical Research Letters}{8}{8}{919-921}.
\newblock
\begin{APACrefURL}
  \url{https://agupubs.onlinelibrary.wiley.com/doi/abs/10.1029/GL008i008p00919}
  \end{APACrefURL}
\newblock
\begin{APACrefDOI} \doi{10.1029/GL008i008p00919} \end{APACrefDOI}
\PrintBackRefs{\CurrentBib}

\bibitem [\protect \citeauthoryear {%
Cane%
}{%
Cane%
}{%
{\protect \APACyear {1979}}%
}]{%
Cane79}
\APACinsertmetastar {%
Cane79}%
\begin{APACrefauthors}%
Cane, H\BPBI V.%
\end{APACrefauthors}%
\unskip\
\newblock
\APACrefYearMonthDay{1979}{12}{}.
\newblock
{\BBOQ}\APACrefatitle {{Spectra of the non-thermal radio radiation from the
  galactic polar regions}} {{Spectra of the non-thermal radio radiation from
  the galactic polar regions}}.{\BBCQ}
\newblock
\APACjournalVolNumPages{Monthly Notices of the Royal Astronomical
  Society}{189}{3}{465-478}.
\newblock
\begin{APACrefURL} \url{https://dx.doi.org/10.1093/mnras/189.3.465}
  \end{APACrefURL}
\newblock
\begin{APACrefDOI} \doi{10.1093/mnras/189.3.465} \end{APACrefDOI}
\PrintBackRefs{\CurrentBib}

\bibitem [\protect \citeauthoryear {%
{Carr}%
, {Desch}%
\BCBL {}\ \BBA {} {Alexander}%
}{%
{Carr}%
\ \protect \BOthers {.}}{%
{\protect \APACyear {1983}}%
}]{%
Carr83}
\APACinsertmetastar {%
Carr83}%
\begin{APACrefauthors}%
{Carr}, T\BPBI D.%
, {Desch}, M\BPBI D.%
\BCBL {}\ \BBA {} {Alexander}, J\BPBI K.%
\end{APACrefauthors}%
\unskip\
\newblock
\APACrefYearMonthDay{1983}{}{}.
\newblock
{\BBOQ}\APACrefatitle {{Phenomenology of magnetospheric radio emissions}}
  {{Phenomenology of magnetospheric radio emissions}}.{\BBCQ}
\newblock
\BIn{} A\BPBI J.~{Dessler}\ (\BED), \APACrefbtitle {Physics of the Jovian
  Magnetosphere} {Physics of the jovian magnetosphere}\ (\BPG~226-284).
\PrintBackRefs{\CurrentBib}

\bibitem [\protect \citeauthoryear {%
Chandran%
, Renuka%
\BCBL {}\ \BBA {} Venugopal%
}{%
Chandran%
\ \protect \BOthers {.}}{%
{\protect \APACyear {2013}}%
}]{%
Chandran13}
\APACinsertmetastar {%
Chandran13}%
\begin{APACrefauthors}%
Chandran, S\BPBI R.%
, Renuka, G.%
\BCBL {}\ \BBA {} Venugopal, C.%
\end{APACrefauthors}%
\unskip\
\newblock
\APACrefYearMonthDay{2013}{}{}.
\newblock
{\BBOQ}\APACrefatitle {Plasma electron temperature variability in lunar surface
  potential and in electric field under average solar wind conditions} {Plasma
  electron temperature variability in lunar surface potential and in electric
  field under average solar wind conditions}.{\BBCQ}
\newblock
\APACjournalVolNumPages{Advances in Space Research}{51}{9}{1622 - 1626}.
\newblock
\begin{APACrefURL}
  \url{http://www.sciencedirect.com/science/article/pii/S0273117713000380}
  \end{APACrefURL}
\newblock
\begin{APACrefDOI} \doi{https://doi.org/10.1016/j.asr.2013.01.016}
  \end{APACrefDOI}
\PrintBackRefs{\CurrentBib}

\bibitem [\protect \citeauthoryear {%
Chin%
\ \protect \BOthers {.}}{%
Chin%
\ \protect \BOthers {.}}{%
{\protect \APACyear {2007}}%
}]{%
Chin2007}
\APACinsertmetastar {%
Chin2007}%
\begin{APACrefauthors}%
Chin, G.%
, Brylow, S.%
, Foote, M.%
, Garvin, J.%
, Kasper, J.%
, Keller, J.%
\BDBL {}Zuber, M.%
\end{APACrefauthors}%
\unskip\
\newblock
\APACrefYearMonthDay{2007}{Apr}{01}.
\newblock
{\BBOQ}\APACrefatitle {Lunar Reconnaissance Orbiter Overview: The Instrument
  Suite and Mission} {Lunar reconnaissance orbiter overview: The instrument
  suite and mission}.{\BBCQ}
\newblock
\APACjournalVolNumPages{Space Science Reviews}{129}{4}{391--419}.
\newblock
\begin{APACrefURL} \url{https://doi.org/10.1007/s11214-007-9153-y}
  \end{APACrefURL}
\newblock
\begin{APACrefDOI} \doi{10.1007/s11214-007-9153-y} \end{APACrefDOI}
\PrintBackRefs{\CurrentBib}

\bibitem [\protect \citeauthoryear {%
Colwell%
, Batiste%
, Horányi%
, Robertson%
\BCBL {}\ \BBA {} Sture%
}{%
Colwell%
\ \protect \BOthers {.}}{%
{\protect \APACyear {2007}}%
}]{%
Colwell07}
\APACinsertmetastar {%
Colwell07}%
\begin{APACrefauthors}%
Colwell, J\BPBI E.%
, Batiste, S.%
, Horányi, M.%
, Robertson, S.%
\BCBL {}\ \BBA {} Sture, S.%
\end{APACrefauthors}%
\unskip\
\newblock
\APACrefYearMonthDay{2007}{}{}.
\newblock
{\BBOQ}\APACrefatitle {Lunar surface: Dust dynamics and regolith mechanics}
  {Lunar surface: Dust dynamics and regolith mechanics}.{\BBCQ}
\newblock
\APACjournalVolNumPages{Reviews of Geophysics}{45}{2}{}.
\newblock
\begin{APACrefURL}
  \url{https://agupubs.onlinelibrary.wiley.com/doi/abs/10.1029/2005RG000184}
  \end{APACrefURL}
\newblock
\begin{APACrefDOI} \doi{10.1029/2005RG000184} \end{APACrefDOI}
\PrintBackRefs{\CurrentBib}

\bibitem [\protect \citeauthoryear {%
{Cornwell}%
}{%
{Cornwell}%
}{%
{\protect \APACyear {2008}}%
}]{%
Cornwell2008}
\APACinsertmetastar {%
Cornwell2008}%
\begin{APACrefauthors}%
{Cornwell}, T\BPBI J.%
\end{APACrefauthors}%
\unskip\
\newblock
\APACrefYearMonthDay{2008}{Oct}{}.
\newblock
{\BBOQ}\APACrefatitle {Multiscale CLEAN Deconvolution of Radio Synthesis
  Images} {Multiscale clean deconvolution of radio synthesis images}.{\BBCQ}
\newblock
\APACjournalVolNumPages{IEEE Journal of Selected Topics in Signal
  Processing}{2}{5}{793-801}.
\newblock
\begin{APACrefDOI} \doi{10.1109/JSTSP.2008.2006388} \end{APACrefDOI}
\PrintBackRefs{\CurrentBib}

\bibitem [\protect \citeauthoryear {%
{Ellingson}%
}{%
{Ellingson}%
}{%
{\protect \APACyear {2011}}%
}]{%
Ellingson11}
\APACinsertmetastar {%
Ellingson11}%
\begin{APACrefauthors}%
{Ellingson}, S\BPBI W.%
\end{APACrefauthors}%
\unskip\
\newblock
\APACrefYearMonthDay{2011}{June}{}.
\newblock
{\BBOQ}\APACrefatitle {Sensitivity of Antenna Arrays for Long-Wavelength Radio
  Astronomy} {Sensitivity of antenna arrays for long-wavelength radio
  astronomy}.{\BBCQ}
\newblock
\APACjournalVolNumPages{IEEE Transactions on Antennas and
  Propagation}{59}{6}{1855-1863}.
\newblock
\begin{APACrefDOI} \doi{10.1109/TAP.2011.2122230} \end{APACrefDOI}
\PrintBackRefs{\CurrentBib}

\bibitem [\protect \citeauthoryear {%
{Ellingson}%
\ \protect \BOthers {.}}{%
{Ellingson}%
\ \protect \BOthers {.}}{%
{\protect \APACyear {2009}}%
}]{%
LWA2009}
\APACinsertmetastar {%
LWA2009}%
\begin{APACrefauthors}%
{Ellingson}, S\BPBI W.%
, {Clarke}, T\BPBI E.%
, {Cohen}, A.%
, {Craig}, J.%
, {Kassim}, N\BPBI E.%
, {Pihlstrom}, Y.%
\BDBL {}{Taylor}, G\BPBI B.%
\end{APACrefauthors}%
\unskip\
\newblock
\APACrefYearMonthDay{2009}{Aug}{}.
\newblock
{\BBOQ}\APACrefatitle {The Long Wavelength Array} {The long wavelength
  array}.{\BBCQ}
\newblock
\APACjournalVolNumPages{Proceedings of the IEEE}{97}{8}{1421-1430}.
\newblock
\begin{APACrefDOI} \doi{10.1109/JPROC.2009.2015683} \end{APACrefDOI}
\PrintBackRefs{\CurrentBib}

\bibitem [\protect \citeauthoryear {%
Elsmore%
}{%
Elsmore%
}{%
{\protect \APACyear {1957}}%
}]{%
Elsmore57}
\APACinsertmetastar {%
Elsmore57}%
\begin{APACrefauthors}%
Elsmore, B.%
\end{APACrefauthors}%
\unskip\
\newblock
\APACrefYearMonthDay{1957}{}{}.
\newblock
{\BBOQ}\APACrefatitle {Radio observations of the lunar atmosphere} {Radio
  observations of the lunar atmosphere}.{\BBCQ}
\newblock
\APACjournalVolNumPages{The Philosophical Magazine: A Journal of Theoretical
  Experimental and Applied Physics}{2}{20}{1040-1046}.
\newblock
\begin{APACrefURL} \url{https://doi.org/10.1080/14786435708238210}
  \end{APACrefURL}
\newblock
\begin{APACrefDOI} \doi{10.1080/14786435708238210} \end{APACrefDOI}
\PrintBackRefs{\CurrentBib}

\bibitem [\protect \citeauthoryear {%
Foster%
\ \protect \BOthers {.}}{%
Foster%
\ \protect \BOthers {.}}{%
{\protect \APACyear {2016}}%
}]{%
Foster16}
\APACinsertmetastar {%
Foster16}%
\begin{APACrefauthors}%
Foster, J\BPBI C.%
, Erickson, P\BPBI J.%
, Baker, D\BPBI N.%
, Jaynes, A\BPBI N.%
, Mishin, E\BPBI V.%
, Fennel, J\BPBI F.%
\BDBL {}Kanekal, S\BPBI G.%
\end{APACrefauthors}%
\unskip\
\newblock
\APACrefYearMonthDay{2016}{}{}.
\newblock
{\BBOQ}\APACrefatitle {Observations of the impenetrable barrier, the
  plasmapause, and the VLF bubble during the 17 March 2015 storm} {Observations
  of the impenetrable barrier, the plasmapause, and the vlf bubble during the
  17 march 2015 storm}.{\BBCQ}
\newblock
\APACjournalVolNumPages{Journal of Geophysical Research: Space
  Physics}{121}{6}{5537-5548}.
\newblock
\begin{APACrefURL}
  \url{https://agupubs.onlinelibrary.wiley.com/doi/abs/10.1002/2016JA022509}
  \end{APACrefURL}
\newblock
\begin{APACrefDOI} \doi{10.1002/2016JA022509} \end{APACrefDOI}
\PrintBackRefs{\CurrentBib}

\bibitem [\protect \citeauthoryear {%
{Girard}%
}{%
{Girard}%
}{%
{\protect \APACyear {2013}}%
}]{%
Girard13thesis}
\APACinsertmetastar {%
Girard13thesis}%
\begin{APACrefauthors}%
{Girard}, J.%
\end{APACrefauthors}%
\unskip\
\newblock
\APACrefYearMonthDay{2013}{}{}.
\newblock
{\BBOQ}\APACrefatitle {{D\`{e}veloppement de la Super Station LOFAR \&
  observations plan\`{e}taires avec LOFAR}} {{D\`{e}veloppement de la Super
  Station LOFAR \& observations plan\`{e}taires avec LOFAR}}.{\BBCQ}
\newblock
\APACjournalVolNumPages{Instrumentation et m\`{e}thodes pour l'astrophysique
  [astro-ph.IM]. Th\`{e}se de Doctorat, Observatoire de Paris, France}{}{}{}.
\newblock
\begin{APACrefURL} \url{tel-00835834v2} \end{APACrefURL}
\PrintBackRefs{\CurrentBib}

\bibitem [\protect \citeauthoryear {%
Girard%
\ \protect \BOthers {.}}{%
Girard%
\ \protect \BOthers {.}}{%
{\protect \APACyear {2016}}%
}]{%
Girard:2016gy}
\APACinsertmetastar {%
Girard:2016gy}%
\begin{APACrefauthors}%
Girard, J\BPBI N.%
, Zarka, P.%
, Tasse, C.%
, Hess, S.%
, de Pater, I.%
, Santos-Costa, D.%
\BDBL {}Wucknitz, O.%
\end{APACrefauthors}%
\unskip\
\newblock
\APACrefYearMonthDay{2016}{{\APACmonth{02}}}{}.
\newblock
{\BBOQ}\APACrefatitle {{Imaging Jupiter's radiation belts down to 127 MHz with
  LOFAR}} {{Imaging Jupiter's radiation belts down to 127 MHz with
  LOFAR}}.{\BBCQ}
\newblock
\APACjournalVolNumPages{Astronomy and Astrophysics}{587}{}{A3}.
\PrintBackRefs{\CurrentBib}

\bibitem [\protect \citeauthoryear {%
Graham%
\ \BBA {} Reckart%
}{%
Graham%
\ \BBA {} Reckart%
}{%
{\protect \APACyear {2019}}%
}]{%
NASACLPS}
\APACinsertmetastar {%
NASACLPS}%
\begin{APACrefauthors}%
Graham, S.%
\BCBT {}\ \BBA {} Reckart, T.%
\end{APACrefauthors}%
\unskip\
\newblock
\APACrefYearMonthDay{2019}{Mar}{}.
\newblock
\APACrefbtitle {NASA-provided Lunar Payloads.} {Nasa-provided lunar payloads.}
\newblock
\APACaddressPublisher{}{NASA Glenn Research Center}.
\newblock
\begin{APACrefURL}
  \url{https://www1.grc.nasa.gov/space/planetary-exploration-science-technology-office-pesto/management/nasa-provided-lunar-payloads/}
  \end{APACrefURL}
\PrintBackRefs{\CurrentBib}

\bibitem [\protect \citeauthoryear {%
Gurnett%
}{%
Gurnett%
}{%
{\protect \APACyear {1974}}%
}]{%
Gurnett74}
\APACinsertmetastar {%
Gurnett74}%
\begin{APACrefauthors}%
Gurnett, D\BPBI A.%
\end{APACrefauthors}%
\unskip\
\newblock
\APACrefYearMonthDay{1974}{}{}.
\newblock
{\BBOQ}\APACrefatitle {The Earth as a radio source: Terrestrial kilometric
  radiation} {The earth as a radio source: Terrestrial kilometric
  radiation}.{\BBCQ}
\newblock
\APACjournalVolNumPages{Journal of Geophysical Research
  (1896-1977)}{79}{28}{4227-4238}.
\newblock
\begin{APACrefURL}
  \url{https://agupubs.onlinelibrary.wiley.com/doi/abs/10.1029/JA079i028p04227}
  \end{APACrefURL}
\newblock
\begin{APACrefDOI} \doi{10.1029/JA079i028p04227} \end{APACrefDOI}
\PrintBackRefs{\CurrentBib}

\bibitem [\protect \citeauthoryear {%
Han%
\ \protect \BOthers {.}}{%
Han%
\ \protect \BOthers {.}}{%
{\protect \APACyear {2018}}%
}]{%
Han18}
\APACinsertmetastar {%
Han18}%
\begin{APACrefauthors}%
Han, S.%
, Murakami, G.%
, Kita, H.%
, Tsuchiya, F.%
, Tao, C.%
, Misawa, H.%
\BDBL {}Nakamura, M.%
\end{APACrefauthors}%
\unskip\
\newblock
\APACrefYearMonthDay{2018}{}{}.
\newblock
{\BBOQ}\APACrefatitle {Investigating Solar Wind-Driven Electric Field Influence
  on Long-Term Dynamics of Jovian Synchrotron Radiation} {Investigating solar
  wind-driven electric field influence on long-term dynamics of jovian
  synchrotron radiation}.{\BBCQ}
\newblock
\APACjournalVolNumPages{Journal of Geophysical Research: Space
  Physics}{123}{11}{9508-9516}.
\newblock
\begin{APACrefURL}
  \url{https://agupubs.onlinelibrary.wiley.com/doi/abs/10.1029/2018JA025849}
  \end{APACrefURL}
\newblock
\begin{APACrefDOI} \doi{10.1029/2018JA025849} \end{APACrefDOI}
\PrintBackRefs{\CurrentBib}

\bibitem [\protect \citeauthoryear {%
Hicks%
\ \protect \BOthers {.}}{%
Hicks%
\ \protect \BOthers {.}}{%
{\protect \APACyear {2012}}%
}]{%
Hicks_2012}
\APACinsertmetastar {%
Hicks_2012}%
\begin{APACrefauthors}%
Hicks, B\BPBI C.%
, Paravastu-Dalal, N.%
, Stewart, K\BPBI P.%
, Erickson, W\BPBI C.%
, Ray, P\BPBI S.%
, Kassim, N\BPBI E.%
\BDBL {}Weiler, K\BPBI W.%
\end{APACrefauthors}%
\unskip\
\newblock
\APACrefYearMonthDay{2012}{oct}{}.
\newblock
{\BBOQ}\APACrefatitle {A Wide-Band, Active Antenna System for Long Wavelength
  Radio Astronomy} {A wide-band, active antenna system for long wavelength
  radio astronomy}.{\BBCQ}
\newblock
\APACjournalVolNumPages{Publications of the Astronomical Society of the
  Pacific}{124}{920}{1090--1104}.
\newblock
\begin{APACrefURL} \url{https://doi.org/10.1086\%2F668121} \end{APACrefURL}
\newblock
\begin{APACrefDOI} \doi{10.1086/668121} \end{APACrefDOI}
\PrintBackRefs{\CurrentBib}

\bibitem [\protect \citeauthoryear {%
{H{\"o}gbom}%
}{%
{H{\"o}gbom}%
}{%
{\protect \APACyear {1974}}%
}]{%
hogbom74}
\APACinsertmetastar {%
hogbom74}%
\begin{APACrefauthors}%
{H{\"o}gbom}, J\BPBI A.%
\end{APACrefauthors}%
\unskip\
\newblock
\APACrefYearMonthDay{1974}{{\APACmonth{06}}}{}.
\newblock
{\BBOQ}\APACrefatitle {{Aperture Synthesis with a Non-Regular Distribution of
  Interferometer Baselines}} {{Aperture Synthesis with a Non-Regular
  Distribution of Interferometer Baselines}}.{\BBCQ}
\newblock
\APACjournalVolNumPages{Astronomy and Astrophysics Supplement}{15}{}{417}.
\PrintBackRefs{\CurrentBib}

\bibitem [\protect \citeauthoryear {%
Hughes%
\ \BBA {} LaBelle%
}{%
Hughes%
\ \BBA {} LaBelle%
}{%
{\protect \APACyear {1998}}%
}]{%
Hughes98}
\APACinsertmetastar {%
Hughes98}%
\begin{APACrefauthors}%
Hughes, J\BPBI M.%
\BCBT {}\ \BBA {} LaBelle, J.%
\end{APACrefauthors}%
\unskip\
\newblock
\APACrefYearMonthDay{1998}{}{}.
\newblock
{\BBOQ}\APACrefatitle {The latitude dependence of auroral roar} {The latitude
  dependence of auroral roar}.{\BBCQ}
\newblock
\APACjournalVolNumPages{Journal of Geophysical Research: Space
  Physics}{103}{A7}{14911-14915}.
\newblock
\begin{APACrefURL}
  \url{https://agupubs.onlinelibrary.wiley.com/doi/abs/10.1029/98JA01038}
  \end{APACrefURL}
\newblock
\begin{APACrefDOI} \doi{10.1029/98JA01038} \end{APACrefDOI}
\PrintBackRefs{\CurrentBib}

\bibitem [\protect \citeauthoryear {%
Imamura%
\ \protect \BOthers {.}}{%
Imamura%
\ \protect \BOthers {.}}{%
{\protect \APACyear {2012}}%
}]{%
Imamura12}
\APACinsertmetastar {%
Imamura12}%
\begin{APACrefauthors}%
Imamura, T.%
, Nabatov, A.%
, Mochizuki, N.%
, Iwata, T.%
, Hanada, H.%
, Matsumoto, K.%
\BDBL {}Saito, A.%
\end{APACrefauthors}%
\unskip\
\newblock
\APACrefYearMonthDay{2012}{}{}.
\newblock
{\BBOQ}\APACrefatitle {Radio occultation measurement of the electron density
  near the lunar surface using a subsatellite on the SELENE mission} {Radio
  occultation measurement of the electron density near the lunar surface using
  a subsatellite on the selene mission}.{\BBCQ}
\newblock
\APACjournalVolNumPages{Journal of Geophysical Research: Space
  Physics}{117}{A6}{}.
\newblock
\begin{APACrefURL}
  \url{https://agupubs.onlinelibrary.wiley.com/doi/abs/10.1029/2011JA017293}
  \end{APACrefURL}
\newblock
\begin{APACrefDOI} \doi{10.1029/2011JA017293} \end{APACrefDOI}
\PrintBackRefs{\CurrentBib}

\bibitem [\protect \citeauthoryear {%
Jun%
\ \BBA {} Garrett%
}{%
Jun%
\ \BBA {} Garrett%
}{%
{\protect \APACyear {2005}}%
}]{%
Jun05}
\APACinsertmetastar {%
Jun05}%
\begin{APACrefauthors}%
Jun, I.%
\BCBT {}\ \BBA {} Garrett, H\BPBI B.%
\end{APACrefauthors}%
\unskip\
\newblock
\APACrefYearMonthDay{2005}{12}{}.
\newblock
{\BBOQ}\APACrefatitle {{Comparison of high-energy trapped particle environments
  at the earth and jupiter}} {{Comparison of high-energy trapped particle
  environments at the earth and jupiter}}.{\BBCQ}
\newblock
\APACjournalVolNumPages{Radiation Protection Dosimetry}{116}{1-4}{50-54}.
\newblock
\begin{APACrefURL} \url{https://doi.org/10.1093/rpd/nci074} \end{APACrefURL}
\newblock
\begin{APACrefDOI} \doi{10.1093/rpd/nci074} \end{APACrefDOI}
\PrintBackRefs{\CurrentBib}

\bibitem [\protect \citeauthoryear {%
KETO%
}{%
KETO%
}{%
{\protect \APACyear {2012}}%
}]{%
Keto2012}
\APACinsertmetastar {%
Keto2012}%
\begin{APACrefauthors}%
KETO, E.%
\end{APACrefauthors}%
\unskip\
\newblock
\APACrefYearMonthDay{2012}{}{}.
\newblock
{\BBOQ}\APACrefatitle {HIERARCHICAL CONFIGURATIONS FOR CROSS-CORRELATION
  INTERFEROMETERS WITH MANY ELEMENTS} {Hierarchical configurations for
  cross-correlation interferometers with many elements}.{\BBCQ}
\newblock
\APACjournalVolNumPages{Journal of Astronomical
  Instrumentation}{01}{01}{1250007}.
\newblock
\begin{APACrefURL} \url{https://doi.org/10.1142/S2251171712500079}
  \end{APACrefURL}
\newblock
\begin{APACrefDOI} \doi{10.1142/S2251171712500079} \end{APACrefDOI}
\PrintBackRefs{\CurrentBib}

\bibitem [\protect \citeauthoryear {%
{LaBelle}%
, {Shepherd}%
\BCBL {}\ \BBA {} {Trimpi}%
}{%
{LaBelle}%
\ \protect \BOthers {.}}{%
{\protect \APACyear {1997}}%
}]{%
LaBelle97}
\APACinsertmetastar {%
LaBelle97}%
\begin{APACrefauthors}%
{LaBelle}, J.%
, {Shepherd}, S\BPBI G.%
\BCBL {}\ \BBA {} {Trimpi}, M\BPBI L.%
\end{APACrefauthors}%
\unskip\
\newblock
\APACrefYearMonthDay{1997}{{\APACmonth{09}}}{}.
\newblock
{\BBOQ}\APACrefatitle {{Observations of auroral medium frequency bursts}}
  {{Observations of auroral medium frequency bursts}}.{\BBCQ}
\newblock
\APACjournalVolNumPages{Journal of Geophysical Research}{102}{}{22221-22232}.
\newblock
\begin{APACrefDOI} \doi{10.1029/97JA01905} \end{APACrefDOI}
\PrintBackRefs{\CurrentBib}

\bibitem [\protect \citeauthoryear {%
LaBelle%
, Trimpi%
, Brittain%
\BCBL {}\ \BBA {} Weatherwax%
}{%
LaBelle%
\ \protect \BOthers {.}}{%
{\protect \APACyear {1995}}%
}]{%
LaBelle95}
\APACinsertmetastar {%
LaBelle95}%
\begin{APACrefauthors}%
LaBelle, J.%
, Trimpi, M\BPBI L.%
, Brittain, R.%
\BCBL {}\ \BBA {} Weatherwax, A\BPBI T.%
\end{APACrefauthors}%
\unskip\
\newblock
\APACrefYearMonthDay{1995}{}{}.
\newblock
{\BBOQ}\APACrefatitle {Fine structure of auroral roar emissions} {Fine
  structure of auroral roar emissions}.{\BBCQ}
\newblock
\APACjournalVolNumPages{Journal of Geophysical Research: Space
  Physics}{100}{A11}{21953-21959}.
\newblock
\begin{APACrefURL}
  \url{https://agupubs.onlinelibrary.wiley.com/doi/abs/10.1029/95JA01551}
  \end{APACrefURL}
\newblock
\begin{APACrefDOI} \doi{10.1029/95JA01551} \end{APACrefDOI}
\PrintBackRefs{\CurrentBib}

\bibitem [\protect \citeauthoryear {%
Lamy%
, Zarka%
, Cecconi%
\BCBL {}\ \BBA {} Prang{\'e}%
}{%
Lamy%
\ \protect \BOthers {.}}{%
{\protect \APACyear {2010}}%
}]{%
lamy_GRL_10a}
\APACinsertmetastar {%
lamy_GRL_10a}%
\begin{APACrefauthors}%
Lamy, L.%
, Zarka, P.%
, Cecconi, B.%
\BCBL {}\ \BBA {} Prang{\'e}, R.%
\end{APACrefauthors}%
\unskip\
\newblock
\APACrefYearMonthDay{2010}{}{}.
\newblock
{\BBOQ}\APACrefatitle {{AKR diurnal, semi-diurnal and shorter term modulations
  disentangled by Cassini/RPWS observations}} {{AKR diurnal, semi-diurnal and
  shorter term modulations disentangled by Cassini/RPWS observations}}.{\BBCQ}
\newblock
\APACjournalVolNumPages{J. Geophys. Res.}{115}{A09221}{}.
\PrintBackRefs{\CurrentBib}

\bibitem [\protect \citeauthoryear {%
Maget%
\ \protect \BOthers {.}}{%
Maget%
\ \protect \BOthers {.}}{%
{\protect \APACyear {2015}}%
}]{%
Maget15}
\APACinsertmetastar {%
Maget15}%
\begin{APACrefauthors}%
Maget, V.%
, Sicard-Piet, A.%
, Bourdarie, S.%
, Lazaro, D.%
, Turner, D\BPBI L.%
, Daglis, I\BPBI A.%
\BCBL {}\ \BBA {} Sandberg, I.%
\end{APACrefauthors}%
\unskip\
\newblock
\APACrefYearMonthDay{2015}{}{}.
\newblock
{\BBOQ}\APACrefatitle {Improved outer boundary conditions for outer radiation
  belt data assimilation using THEMIS-SST data and the Salammbo-EnKF code}
  {Improved outer boundary conditions for outer radiation belt data
  assimilation using themis-sst data and the salammbo-enkf code}.{\BBCQ}
\newblock
\APACjournalVolNumPages{Journal of Geophysical Research: Space
  Physics}{120}{7}{5608-5622}.
\newblock
\begin{APACrefURL}
  \url{https://agupubs.onlinelibrary.wiley.com/doi/abs/10.1002/2015JA021001}
  \end{APACrefURL}
\newblock
\begin{APACrefDOI} \doi{10.1002/2015JA021001} \end{APACrefDOI}
\PrintBackRefs{\CurrentBib}

\bibitem [\protect \citeauthoryear {%
{Manning, R.}%
\ \BBA {} {Dulk, G. A.}%
}{%
{Manning, R.}%
\ \BBA {} {Dulk, G. A.}%
}{%
{\protect \APACyear {2001}}%
}]{%
manning01}
\APACinsertmetastar {%
manning01}%
\begin{APACrefauthors}%
{Manning, R.}%
\BCBT {}\ \BBA {} {Dulk, G. A.}%
\end{APACrefauthors}%
\unskip\
\newblock
\APACrefYearMonthDay{2001}{}{}.
\newblock
{\BBOQ}\APACrefatitle {The Galactic background radiation from 0.2 to 13.8 MHz}
  {The galactic background radiation from 0.2 to 13.8 mhz}.{\BBCQ}
\newblock
\APACjournalVolNumPages{Astronomy \& Astrophysics}{372}{2}{663-666}.
\newblock
\begin{APACrefURL} \url{https://doi.org/10.1051/0004-6361:20010516}
  \end{APACrefURL}
\newblock
\begin{APACrefDOI} \doi{10.1051/0004-6361:20010516} \end{APACrefDOI}
\PrintBackRefs{\CurrentBib}

\bibitem [\protect \citeauthoryear {%
{Mart{\'{\i}}-Vidal}%
, {P{\'e}rez-Torres}%
\BCBL {}\ \BBA {} {Lobanov}%
}{%
{Mart{\'{\i}}-Vidal}%
\ \protect \BOthers {.}}{%
{\protect \APACyear {2012}}%
}]{%
Vidal12}
\APACinsertmetastar {%
Vidal12}%
\begin{APACrefauthors}%
{Mart{\'{\i}}-Vidal}, I.%
, {P{\'e}rez-Torres}, M\BPBI A.%
\BCBL {}\ \BBA {} {Lobanov}, A\BPBI P.%
\end{APACrefauthors}%
\unskip\
\newblock
\APACrefYearMonthDay{2012}{{\APACmonth{05}}}{}.
\newblock
{\BBOQ}\APACrefatitle {{Over-resolution of compact sources in interferometric
  observations}} {{Over-resolution of compact sources in interferometric
  observations}}.{\BBCQ}
\newblock
\APACjournalVolNumPages{Astronomy \& Astrophysics}{541}{}{A135}.
\newblock
\begin{APACrefDOI} \doi{10.1051/0004-6361/201118334} \end{APACrefDOI}
\PrintBackRefs{\CurrentBib}

\bibitem [\protect \citeauthoryear {%
{McMullin}%
, {Waters}%
, {Schiebel}%
, {Young}%
\BCBL {}\ \BBA {} {Golap}%
}{%
{McMullin}%
\ \protect \BOthers {.}}{%
{\protect \APACyear {2007}}%
}]{%
Casa07}
\APACinsertmetastar {%
Casa07}%
\begin{APACrefauthors}%
{McMullin}, J\BPBI P.%
, {Waters}, B.%
, {Schiebel}, D.%
, {Young}, W.%
\BCBL {}\ \BBA {} {Golap}, K.%
\end{APACrefauthors}%
\unskip\
\newblock
\APACrefYearMonthDay{2007}{{\APACmonth{10}}}{}.
\newblock
{\BBOQ}\APACrefatitle {{CASA Architecture and Applications}} {{CASA
  Architecture and Applications}}.{\BBCQ}
\newblock
\BIn{} R\BPBI A.~{Shaw}, F.~{Hill}\BCBL {}\ \BBA {} D\BPBI J.~{Bell}\ (\BEDS),
  \APACrefbtitle {Astronomical Data Analysis Software and Systems XVI}
  {Astronomical data analysis software and systems xvi}\ (\BVOL~376, \BPG~127).
\PrintBackRefs{\CurrentBib}

\bibitem [\protect \citeauthoryear {%
Meyer-Vernet%
, Hoang%
, Issautier%
, Moncuquet%
\BCBL {}\ \BBA {} Marcos%
}{%
Meyer-Vernet%
\ \protect \BOthers {.}}{%
{\protect \APACyear {2000}}%
}]{%
Vernet13}
\APACinsertmetastar {%
Vernet13}%
\begin{APACrefauthors}%
Meyer-Vernet, N.%
, Hoang, S.%
, Issautier, K.%
, Moncuquet, M.%
\BCBL {}\ \BBA {} Marcos, G.%
\end{APACrefauthors}%
\unskip\
\newblock
\APACrefYearMonthDay{2000}{}{}.
\newblock
{\BBOQ}\APACrefatitle {Plasma Thermal Noise: The Long Wavelength Radio Limit}
  {Plasma thermal noise: The long wavelength radio limit}.{\BBCQ}
\newblock
\BIn{} \APACrefbtitle {Radio Astronomy at Long Wavelengths} {Radio astronomy at
  long wavelengths}\ (\BPG~67-74).
\newblock
\APACaddressPublisher{}{American Geophysical Union (AGU)}.
\newblock
\begin{APACrefURL}
  \url{https://agupubs.onlinelibrary.wiley.com/doi/abs/10.1029/GM119p0067}
  \end{APACrefURL}
\newblock
\begin{APACrefDOI} \doi{10.1029/GM119p0067} \end{APACrefDOI}
\PrintBackRefs{\CurrentBib}

\bibitem [\protect \citeauthoryear {%
Meyer-Vernet%
\ \BBA {} Perche%
}{%
Meyer-Vernet%
\ \BBA {} Perche%
}{%
{\protect \APACyear {1989}}%
}]{%
Vernet89}
\APACinsertmetastar {%
Vernet89}%
\begin{APACrefauthors}%
Meyer-Vernet, N.%
\BCBT {}\ \BBA {} Perche, C.%
\end{APACrefauthors}%
\unskip\
\newblock
\APACrefYearMonthDay{1989}{}{}.
\newblock
{\BBOQ}\APACrefatitle {Tool kit for antennae and thermal noise near the plasma
  frequency} {Tool kit for antennae and thermal noise near the plasma
  frequency}.{\BBCQ}
\newblock
\APACjournalVolNumPages{Journal of Geophysical Research: Space
  Physics}{94}{A3}{2405-2415}.
\newblock
\begin{APACrefURL}
  \url{https://agupubs.onlinelibrary.wiley.com/doi/abs/10.1029/JA094iA03p02405}
  \end{APACrefURL}
\newblock
\begin{APACrefDOI} \doi{10.1029/JA094iA03p02405} \end{APACrefDOI}
\PrintBackRefs{\CurrentBib}

\bibitem [\protect \citeauthoryear {%
Mishra%
\ \BBA {} Misra%
}{%
Mishra%
\ \BBA {} Misra%
}{%
{\protect \APACyear {2018}}%
}]{%
Mishra18}
\APACinsertmetastar {%
Mishra18}%
\begin{APACrefauthors}%
Mishra, S\BPBI K.%
\BCBT {}\ \BBA {} Misra, S.%
\end{APACrefauthors}%
\unskip\
\newblock
\APACrefYearMonthDay{2018}{}{}.
\newblock
{\BBOQ}\APACrefatitle {An analytical investigation: Effect of solar wind on
  lunar photoelectron sheath} {An analytical investigation: Effect of solar
  wind on lunar photoelectron sheath}.{\BBCQ}
\newblock
\APACjournalVolNumPages{Physics of Plasmas}{25}{2}{023702}.
\newblock
\begin{APACrefURL} \url{https://doi.org/10.1063/1.5021260} \end{APACrefURL}
\newblock
\begin{APACrefDOI} \doi{10.1063/1.5021260} \end{APACrefDOI}
\PrintBackRefs{\CurrentBib}

\bibitem [\protect \citeauthoryear {%
Morgan%
\ \BBA {} A.~Gurnett%
}{%
Morgan%
\ \BBA {} A.~Gurnett%
}{%
{\protect \APACyear {1991}}%
}]{%
Morgan91}
\APACinsertmetastar {%
Morgan91}%
\begin{APACrefauthors}%
Morgan, D.%
\BCBT {}\ \BBA {} A.~Gurnett, D.%
\end{APACrefauthors}%
\unskip\
\newblock
\APACrefYearMonthDay{1991}{06}{}.
\newblock
{\BBOQ}\APACrefatitle {The source location and beaming of terrestrial continuum
  radiation} {The source location and beaming of terrestrial continuum
  radiation}.{\BBCQ}
\newblock
\APACjournalVolNumPages{Journal of Geophysical Research}{96}{}{9595-9613}.
\newblock
\begin{APACrefDOI} \doi{10.1029/91JA00314} \end{APACrefDOI}
\PrintBackRefs{\CurrentBib}

\bibitem [\protect \citeauthoryear {%
Mutel%
, Christopher%
\BCBL {}\ \BBA {} Pickett%
}{%
Mutel%
\ \protect \BOthers {.}}{%
{\protect \APACyear {2008}}%
}]{%
Mutel08}
\APACinsertmetastar {%
Mutel08}%
\begin{APACrefauthors}%
Mutel, R\BPBI L.%
, Christopher, I\BPBI W.%
\BCBL {}\ \BBA {} Pickett, J\BPBI S.%
\end{APACrefauthors}%
\unskip\
\newblock
\APACrefYearMonthDay{2008}{}{}.
\newblock
{\BBOQ}\APACrefatitle {Cluster multispacecraft determination of AKR angular
  beaming} {Cluster multispacecraft determination of akr angular
  beaming}.{\BBCQ}
\newblock
\APACjournalVolNumPages{Geophysical Research Letters}{35}{7}{}.
\newblock
\begin{APACrefURL}
  \url{https://agupubs.onlinelibrary.wiley.com/doi/abs/10.1029/2008GL033377}
  \end{APACrefURL}
\newblock
\begin{APACrefDOI} \doi{10.1029/2008GL033377} \end{APACrefDOI}
\PrintBackRefs{\CurrentBib}

\bibitem [\protect \citeauthoryear {%
N\`{e}non%
, Sicard%
\BCBL {}\ \BBA {} Bourdarie%
}{%
N\`{e}non%
\ \protect \BOthers {.}}{%
{\protect \APACyear {2017}}%
}]{%
Nenon2017}
\APACinsertmetastar {%
Nenon2017}%
\begin{APACrefauthors}%
N\`{e}non, Q.%
, Sicard, A.%
\BCBL {}\ \BBA {} Bourdarie, S.%
\end{APACrefauthors}%
\unskip\
\newblock
\APACrefYearMonthDay{2017}{}{}.
\newblock
{\BBOQ}\APACrefatitle {A new physical model of the electron radiation belts of
  Jupiter inside Europa's orbit} {A new physical model of the electron
  radiation belts of jupiter inside europa's orbit}.{\BBCQ}
\newblock
\APACjournalVolNumPages{Journal of Geophysical Research: Space
  Physics}{122}{5}{5148-5167}.
\newblock
\begin{APACrefURL}
  \url{https://agupubs.onlinelibrary.wiley.com/doi/abs/10.1002/2017JA023893}
  \end{APACrefURL}
\newblock
\begin{APACrefDOI} \doi{10.1002/2017JA023893} \end{APACrefDOI}
\PrintBackRefs{\CurrentBib}

\bibitem [\protect \citeauthoryear {%
{Noordam}%
}{%
{Noordam}%
}{%
{\protect \APACyear {2004}}%
}]{%
Noordam04}
\APACinsertmetastar {%
Noordam04}%
\begin{APACrefauthors}%
{Noordam}, J\BPBI E.%
\end{APACrefauthors}%
\unskip\
\newblock
\APACrefYearMonthDay{2004}{{\APACmonth{10}}}{}.
\newblock
{\BBOQ}\APACrefatitle {{LOFAR calibration challenges}} {{LOFAR calibration
  challenges}}.{\BBCQ}
\newblock
\BIn{} J\BPBI M.~{Oschmann} Jr.\ (\BED), \APACrefbtitle {Ground-based
  Telescopes} {Ground-based telescopes}\ (\BVOL\ 5489, \BPG~817-825).
\newblock
\begin{APACrefDOI} \doi{10.1117/12.544262} \end{APACrefDOI}
\PrintBackRefs{\CurrentBib}

\bibitem [\protect \citeauthoryear {%
{Novaco}%
\ \BBA {} {Brown}%
}{%
{Novaco}%
\ \BBA {} {Brown}%
}{%
{\protect \APACyear {1978}}%
}]{%
Novaco78}
\APACinsertmetastar {%
Novaco78}%
\begin{APACrefauthors}%
{Novaco}, J\BPBI C.%
\BCBT {}\ \BBA {} {Brown}, L\BPBI W.%
\end{APACrefauthors}%
\unskip\
\newblock
\APACrefYearMonthDay{1978}{{\APACmonth{04}}}{}.
\newblock
{\BBOQ}\APACrefatitle {{Nonthermal galactic emission below 10 megahertz}}
  {{Nonthermal galactic emission below 10 megahertz}}.{\BBCQ}
\newblock
\APACjournalVolNumPages{The Astrophysical Journal}{221}{}{114-123}.
\newblock
\begin{APACrefDOI} \doi{10.1086/156009} \end{APACrefDOI}
\PrintBackRefs{\CurrentBib}

\bibitem [\protect \citeauthoryear {%
Ondoh%
}{%
Ondoh%
}{%
{\protect \APACyear {2013}}%
}]{%
Ondoh91}
\APACinsertmetastar {%
Ondoh91}%
\begin{APACrefauthors}%
Ondoh, T.%
\end{APACrefauthors}%
\unskip\
\newblock
\APACrefYearMonthDay{2013}{}{}.
\newblock
{\BBOQ}\APACrefatitle {Polar Hiss Observed by Isis Satellites} {Polar hiss
  observed by isis satellites}.{\BBCQ}
\newblock
\BIn{} \APACrefbtitle {Magnetospheric Substorms} {Magnetospheric substorms}\
  (\BPG~387-398).
\newblock
\APACaddressPublisher{}{American Geophysical Union (AGU)}.
\newblock
\begin{APACrefURL}
  \url{https://agupubs.onlinelibrary.wiley.com/doi/abs/10.1029/GM064p0387}
  \end{APACrefURL}
\newblock
\begin{APACrefDOI} \doi{10.1029/GM064p0387} \end{APACrefDOI}
\PrintBackRefs{\CurrentBib}

\bibitem [\protect \citeauthoryear {%
{Pacholczyk}%
}{%
{Pacholczyk}%
}{%
{\protect \APACyear {1970}}%
}]{%
Pacho70}
\APACinsertmetastar {%
Pacho70}%
\begin{APACrefauthors}%
{Pacholczyk}, A\BPBI G.%
\end{APACrefauthors}%
\unskip\
\newblock
\APACrefYear{1970}.
\newblock
\APACrefbtitle {{Radio astrophysics. Nonthermal processes in galactic and
  extragalactic sources}} {{Radio astrophysics. Nonthermal processes in
  galactic and extragalactic sources}}.
\PrintBackRefs{\CurrentBib}

\bibitem [\protect \citeauthoryear {%
{Pierrard}%
, {Lopez Rosson}%
\BCBL {}\ \BBA {} {Botek}%
}{%
{Pierrard}%
\ \protect \BOthers {.}}{%
{\protect \APACyear {2019}}%
}]{%
Pierrard2019}
\APACinsertmetastar {%
Pierrard2019}%
\begin{APACrefauthors}%
{Pierrard}, V.%
, {Lopez Rosson}, G.%
\BCBL {}\ \BBA {} {Botek}, E.%
\end{APACrefauthors}%
\unskip\
\newblock
\APACrefYearMonthDay{2019}{Mar}{}.
\newblock
{\BBOQ}\APACrefatitle {{Dynamics of Megaelectron Volt Electrons Observed in the
  Inner Belt by PROBA-V/EPT}} {{Dynamics of Megaelectron Volt Electrons
  Observed in the Inner Belt by PROBA-V/EPT}}.{\BBCQ}
\newblock
\APACjournalVolNumPages{Journal of Geophysical Research (Space
  Physics)}{124}{3}{1651-1659}.
\newblock
\begin{APACrefDOI} \doi{10.1029/2018JA026289} \end{APACrefDOI}
\PrintBackRefs{\CurrentBib}

\bibitem [\protect \citeauthoryear {%
Planck%
}{%
Planck%
}{%
{\protect \APACyear {1914}}%
}]{%
Planck14}
\APACinsertmetastar {%
Planck14}%
\begin{APACrefauthors}%
Planck, M.%
\end{APACrefauthors}%
\unskip\
\newblock
\APACrefYear{1914}.
\newblock
\APACrefbtitle {The Theory of Heat Radiation} {The theory of heat radiation}.
\newblock
\APACaddressPublisher{Philadelphia, PA, USA}{P. Blakiston's Son and Co.}
\newblock
\APACrefnote{Authorized translation by Morton Masius.}
\PrintBackRefs{\CurrentBib}

\bibitem [\protect \citeauthoryear {%
Rayleigh%
}{%
Rayleigh%
}{%
{\protect \APACyear {1879}}%
}]{%
Rayleigh1879}
\APACinsertmetastar {%
Rayleigh1879}%
\begin{APACrefauthors}%
Rayleigh, L.%
\end{APACrefauthors}%
\unskip\
\newblock
\APACrefYearMonthDay{1879}{}{}.
\newblock
{\BBOQ}\APACrefatitle {XXXI. Investigations in optics, with special reference
  to the spectroscope} {Xxxi. investigations in optics, with special reference
  to the spectroscope}.{\BBCQ}
\newblock
\APACjournalVolNumPages{The London, Edinburgh, and Dublin Philosophical
  Magazine and Journal of Science}{8}{49}{261-274}.
\newblock
\begin{APACrefURL} \url{https://doi.org/10.1080/14786447908639684}
  \end{APACrefURL}
\newblock
\begin{APACrefDOI} \doi{10.1080/14786447908639684} \end{APACrefDOI}
\PrintBackRefs{\CurrentBib}

\bibitem [\protect \citeauthoryear {%
Reames%
}{%
Reames%
}{%
{\protect \APACyear {2013}}%
}]{%
Reames2013}
\APACinsertmetastar {%
Reames2013}%
\begin{APACrefauthors}%
Reames, D\BPBI V.%
\end{APACrefauthors}%
\unskip\
\newblock
\APACrefYearMonthDay{2013}{Jun}{01}.
\newblock
{\BBOQ}\APACrefatitle {The Two Sources of Solar Energetic Particles} {The two
  sources of solar energetic particles}.{\BBCQ}
\newblock
\APACjournalVolNumPages{Space Science Reviews}{175}{1}{53--92}.
\newblock
\begin{APACrefURL} \url{https://doi.org/10.1007/s11214-013-9958-9}
  \end{APACrefURL}
\newblock
\begin{APACrefDOI} \doi{10.1007/s11214-013-9958-9} \end{APACrefDOI}
\PrintBackRefs{\CurrentBib}

\bibitem [\protect \citeauthoryear {%
{Santos-Costa}%
\ \BBA {} {Bourdarie}%
}{%
{Santos-Costa}%
\ \BBA {} {Bourdarie}%
}{%
{\protect \APACyear {2001}}%
}]{%
Costa01}
\APACinsertmetastar {%
Costa01}%
\begin{APACrefauthors}%
{Santos-Costa}, D.%
\BCBT {}\ \BBA {} {Bourdarie}, S\BPBI A.%
\end{APACrefauthors}%
\unskip\
\newblock
\APACrefYearMonthDay{2001}{{\APACmonth{03}}}{}.
\newblock
{\BBOQ}\APACrefatitle {{Modeling the inner Jovian electron radiation belt
  including non-equatorial particles}} {{Modeling the inner Jovian electron
  radiation belt including non-equatorial particles}}.{\BBCQ}
\newblock
\APACjournalVolNumPages{Planet. Space Sci}{49}{}{303-312}.
\newblock
\begin{APACrefDOI} \doi{10.1016/S0032-0633(00)00151-3} \end{APACrefDOI}
\PrintBackRefs{\CurrentBib}

\bibitem [\protect \citeauthoryear {%
{Santos-Costa, D.}%
\ \protect \BOthers {.}}{%
{Santos-Costa, D.}%
\ \protect \BOthers {.}}{%
{\protect \APACyear {2014}}%
}]{%
Costa14}
\APACinsertmetastar {%
Costa14}%
\begin{APACrefauthors}%
{Santos-Costa, D.}%
, {de Pater, I.}%
, {Sault, R. J.}%
, {Janssen, M. A.}%
, {Levin, S. M.}%
\BCBL {}\ \BBA {} {Bolton, S. J.}%
\end{APACrefauthors}%
\unskip\
\newblock
\APACrefYearMonthDay{2014}{}{}.
\newblock
{\BBOQ}\APACrefatitle {Multifrequency analysis of the Jovian electron-belt
  radiation during the Cassini flyby of Jupiter} {Multifrequency analysis of
  the jovian electron-belt radiation during the cassini flyby of
  jupiter}.{\BBCQ}
\newblock
\APACjournalVolNumPages{Astronomy \& Astrophysics}{568}{}{A61}.
\newblock
\begin{APACrefURL} \url{https://doi.org/10.1051/0004-6361/201423896}
  \end{APACrefURL}
\newblock
\begin{APACrefDOI} \doi{10.1051/0004-6361/201423896} \end{APACrefDOI}
\PrintBackRefs{\CurrentBib}

\bibitem [\protect \citeauthoryear {%
Santos-Costa%
\ \BBA {} Bolton%
}{%
Santos-Costa%
\ \BBA {} Bolton%
}{%
{\protect \APACyear {2008}}%
}]{%
Costa08}
\APACinsertmetastar {%
Costa08}%
\begin{APACrefauthors}%
Santos-Costa, D.%
\BCBT {}\ \BBA {} Bolton, S\BPBI J.%
\end{APACrefauthors}%
\unskip\
\newblock
\APACrefYearMonthDay{2008}{}{}.
\newblock
{\BBOQ}\APACrefatitle {Discussing the processes constraining the Jovian
  synchrotron radio emission's features} {Discussing the processes constraining
  the jovian synchrotron radio emission's features}.{\BBCQ}
\newblock
\APACjournalVolNumPages{Planetary and Space Science}{56}{3}{326 - 345}.
\newblock
\begin{APACrefURL}
  \url{http://www.sciencedirect.com/science/article/pii/S0032063307002942}
  \end{APACrefURL}
\newblock
\begin{APACrefDOI} \doi{https://doi.org/10.1016/j.pss.2007.09.008}
  \end{APACrefDOI}
\PrintBackRefs{\CurrentBib}

\bibitem [\protect \citeauthoryear {%
Santos-Costa%
, Bolton%
, Sault%
, Thorne%
\BCBL {}\ \BBA {} Levin%
}{%
Santos-Costa%
\ \protect \BOthers {.}}{%
{\protect \APACyear {2011}}%
}]{%
Costa11}
\APACinsertmetastar {%
Costa11}%
\begin{APACrefauthors}%
Santos-Costa, D.%
, Bolton, S\BPBI J.%
, Sault, R\BPBI J.%
, Thorne, R\BPBI M.%
\BCBL {}\ \BBA {} Levin, S\BPBI M.%
\end{APACrefauthors}%
\unskip\
\newblock
\APACrefYearMonthDay{2011}{}{}.
\newblock
{\BBOQ}\APACrefatitle {VLA observations at 6.2 cm of the response of Jupiter's
  electron belt to the July 2009 event} {Vla observations at 6.2 cm of the
  response of jupiter's electron belt to the july 2009 event}.{\BBCQ}
\newblock
\APACjournalVolNumPages{Journal of Geophysical Research: Space
  Physics}{116}{A12}{}.
\newblock
\begin{APACrefURL}
  \url{https://agupubs.onlinelibrary.wiley.com/doi/abs/10.1029/2011JA016921}
  \end{APACrefURL}
\newblock
\begin{APACrefDOI} \doi{10.1029/2011JA016921} \end{APACrefDOI}
\PrintBackRefs{\CurrentBib}

\bibitem [\protect \citeauthoryear {%
Sazhin%
, Bullough%
\BCBL {}\ \BBA {} Hayakawa%
}{%
Sazhin%
\ \protect \BOthers {.}}{%
{\protect \APACyear {1993}}%
}]{%
Sazhin93}
\APACinsertmetastar {%
Sazhin93}%
\begin{APACrefauthors}%
Sazhin, S.%
, Bullough, K.%
\BCBL {}\ \BBA {} Hayakawa, M.%
\end{APACrefauthors}%
\unskip\
\newblock
\APACrefYearMonthDay{1993}{}{}.
\newblock
{\BBOQ}\APACrefatitle {Auroral hiss: a review} {Auroral hiss: a review}.{\BBCQ}
\newblock
\APACjournalVolNumPages{Planetary and Space Science}{41}{2}{153 - 166}.
\newblock
\begin{APACrefURL}
  \url{http://www.sciencedirect.com/science/article/pii/0032063393900454}
  \end{APACrefURL}
\newblock
\begin{APACrefDOI} \doi{https://doi.org/10.1016/0032-0633(93)90045-4}
  \end{APACrefDOI}
\PrintBackRefs{\CurrentBib}

\bibitem [\protect \citeauthoryear {%
Sicard%
\ \BBA {} Bourdarie%
}{%
Sicard%
\ \BBA {} Bourdarie%
}{%
{\protect \APACyear {2004}}%
}]{%
Sicard04}
\APACinsertmetastar {%
Sicard04}%
\begin{APACrefauthors}%
Sicard, A.%
\BCBT {}\ \BBA {} Bourdarie, S.%
\end{APACrefauthors}%
\unskip\
\newblock
\APACrefYearMonthDay{2004}{}{}.
\newblock
{\BBOQ}\APACrefatitle {Physical Electron Belt Model from Jupiter's surface to
  the orbit of Europa} {Physical electron belt model from jupiter's surface to
  the orbit of europa}.{\BBCQ}
\newblock
\APACjournalVolNumPages{Journal of Geophysical Research: Space
  Physics}{109}{A2}{}.
\newblock
\begin{APACrefURL}
  \url{https://agupubs.onlinelibrary.wiley.com/doi/abs/10.1029/2003JA010203}
  \end{APACrefURL}
\newblock
\begin{APACrefDOI} \doi{10.1029/2003JA010203} \end{APACrefDOI}
\PrintBackRefs{\CurrentBib}

\bibitem [\protect \citeauthoryear {%
Sodha%
\ \BBA {} Mishra%
}{%
Sodha%
\ \BBA {} Mishra%
}{%
{\protect \APACyear {2014}}%
}]{%
Sodha14}
\APACinsertmetastar {%
Sodha14}%
\begin{APACrefauthors}%
Sodha, M\BPBI S.%
\BCBT {}\ \BBA {} Mishra, S\BPBI K.%
\end{APACrefauthors}%
\unskip\
\newblock
\APACrefYearMonthDay{2014}{}{}.
\newblock
{\BBOQ}\APACrefatitle {Lunar photoelectron sheath and levitation of dust}
  {Lunar photoelectron sheath and levitation of dust}.{\BBCQ}
\newblock
\APACjournalVolNumPages{Physics of Plasmas}{21}{9}{093704}.
\newblock
\begin{APACrefURL} \url{https://doi.org/10.1063/1.4896345} \end{APACrefURL}
\newblock
\begin{APACrefDOI} \doi{10.1063/1.4896345} \end{APACrefDOI}
\PrintBackRefs{\CurrentBib}

\bibitem [\protect \citeauthoryear {%
Stubbs%
\ \protect \BOthers {.}}{%
Stubbs%
\ \protect \BOthers {.}}{%
{\protect \APACyear {2011}}%
}]{%
Stubbs11}
\APACinsertmetastar {%
Stubbs11}%
\begin{APACrefauthors}%
Stubbs, T.%
, Glenar, D.%
, Farrell, W.%
, Vondrak, R.%
, Collier, M.%
, Halekas, J.%
\BCBL {}\ \BBA {} Delory, G.%
\end{APACrefauthors}%
\unskip\
\newblock
\APACrefYearMonthDay{2011}{}{}.
\newblock
{\BBOQ}\APACrefatitle {On the role of dust in the lunar ionosphere} {On the
  role of dust in the lunar ionosphere}.{\BBCQ}
\newblock
\APACjournalVolNumPages{Planetary and Space Science}{59}{13}{1659 - 1664}.
\newblock
\begin{APACrefURL}
  \url{http://www.sciencedirect.com/science/article/pii/S0032063311001693}
  \end{APACrefURL}
\newblock
\APACrefnote{Exploring Phobos}
\newblock
\begin{APACrefDOI} \doi{https://doi.org/10.1016/j.pss.2011.05.011}
  \end{APACrefDOI}
\PrintBackRefs{\CurrentBib}

\bibitem [\protect \citeauthoryear {%
{Taylor}%
, {Carilli}%
\BCBL {}\ \BBA {} {Perley}%
}{%
{Taylor}%
\ \protect \BOthers {.}}{%
{\protect \APACyear {1999}}%
}]{%
Taylor99}
\APACinsertmetastar {%
Taylor99}%
\begin{APACrefauthors}%
{Taylor}, G\BPBI B.%
, {Carilli}, C\BPBI L.%
\BCBL {}\ \BBA {} {Perley}, R\BPBI A.%
\end{APACrefauthors}%
\ (\BEDS).
\unskip\
\newblock
\APACrefYearMonthDay{1999}{}{}.
\newblock
\APACrefbtitle {{Synthesis Imaging in Radio Astronomy II}} {{Synthesis Imaging
  in Radio Astronomy II}}\ (\BVOL~180).
\PrintBackRefs{\CurrentBib}

\bibitem [\protect \citeauthoryear {%
Thompson%
, Emerson%
\BCBL {}\ \BBA {} Schwab%
}{%
Thompson%
\ \protect \BOthers {.}}{%
{\protect \APACyear {2007}}%
}]{%
Thompson07}
\APACinsertmetastar {%
Thompson07}%
\begin{APACrefauthors}%
Thompson, A\BPBI R.%
, Emerson, D\BPBI T.%
\BCBL {}\ \BBA {} Schwab, F\BPBI R.%
\end{APACrefauthors}%
\unskip\
\newblock
\APACrefYearMonthDay{2007}{}{}.
\newblock
{\BBOQ}\APACrefatitle {Convenient formulas for quantization efficiency}
  {Convenient formulas for quantization efficiency}.{\BBCQ}
\newblock
\APACjournalVolNumPages{Radio Science}{42}{3}{}.
\newblock
\begin{APACrefURL}
  \url{https://agupubs.onlinelibrary.wiley.com/doi/abs/10.1029/2006RS003585}
  \end{APACrefURL}
\newblock
\begin{APACrefDOI} \doi{10.1029/2006RS003585} \end{APACrefDOI}
\PrintBackRefs{\CurrentBib}

\bibitem [\protect \citeauthoryear {%
{Thompson}%
, {Moran}%
\BCBL {}\ \BBA {} {Swenson}%
}{%
{Thompson}%
\ \protect \BOthers {.}}{%
{\protect \APACyear {1986}}%
}]{%
Thompson86}
\APACinsertmetastar {%
Thompson86}%
\begin{APACrefauthors}%
{Thompson}, A\BPBI R.%
, {Moran}, J\BPBI M.%
\BCBL {}\ \BBA {} {Swenson}, G\BPBI W.%
\end{APACrefauthors}%
\unskip\
\newblock
\APACrefYear{1986}.
\newblock
\APACrefbtitle {{Interferometry and synthesis in radio astronomy}}
  {{Interferometry and synthesis in radio astronomy}}.
\PrintBackRefs{\CurrentBib}

\bibitem [\protect \citeauthoryear {%
{Tingay}%
\ \protect \BOthers {.}}{%
{Tingay}%
\ \protect \BOthers {.}}{%
{\protect \APACyear {2013}}%
}]{%
MWA13}
\APACinsertmetastar {%
MWA13}%
\begin{APACrefauthors}%
{Tingay}, S\BPBI J.%
, {Goeke}, R.%
, {Bowman}, J\BPBI D.%
, {Emrich}, D.%
, {Ord}, S\BPBI M.%
, {Mitchell}, D\BPBI A.%
\BDBL {}{Wyithe}, J\BPBI S\BPBI B.%
\end{APACrefauthors}%
\unskip\
\newblock
\APACrefYearMonthDay{2013}{{\APACmonth{01}}}{}.
\newblock
{\BBOQ}\APACrefatitle {{The Murchison Widefield Array: The Square Kilometre
  Array Precursor at Low Radio Frequencies}} {{The Murchison Widefield Array:
  The Square Kilometre Array Precursor at Low Radio Frequencies}}.{\BBCQ}
\newblock
\APACjournalVolNumPages{Publications of the Astronomical Society of
  Australia}{30}{}{e007}.
\newblock
\begin{APACrefDOI} \doi{10.1017/pasa.2012.007} \end{APACrefDOI}
\PrintBackRefs{\CurrentBib}

\bibitem [\protect \citeauthoryear {%
{van Haarlem, M. P.}%
\ \protect \BOthers {.}}{%
{van Haarlem, M. P.}%
\ \protect \BOthers {.}}{%
{\protect \APACyear {2013}}%
}]{%
LOFAR2013}
\APACinsertmetastar {%
LOFAR2013}%
\begin{APACrefauthors}%
{van Haarlem, M. P.}%
, {Wise, M. W.}%
, {Gunst, A. W.}%
, {Heald, G.}%
, {McKean, J. P.}%
, {Hessels, J. W. T.}%
\BDBL {}{van Zwieten, J.}%
\end{APACrefauthors}%
\unskip\
\newblock
\APACrefYearMonthDay{2013}{}{}.
\newblock
{\BBOQ}\APACrefatitle {LOFAR: The LOw-Frequency ARray} {Lofar: The
  low-frequency array}.{\BBCQ}
\newblock
\APACjournalVolNumPages{A\&A}{556}{}{A2}.
\newblock
\begin{APACrefURL} \url{https://doi.org/10.1051/0004-6361/201220873}
  \end{APACrefURL}
\newblock
\begin{APACrefDOI} \doi{10.1051/0004-6361/201220873} \end{APACrefDOI}
\PrintBackRefs{\CurrentBib}

\bibitem [\protect \citeauthoryear {%
{Vasil'Ev}%
\ \protect \BOthers {.}}{%
{Vasil'Ev}%
\ \protect \BOthers {.}}{%
{\protect \APACyear {1974}}%
}]{%
Vasil74}
\APACinsertmetastar {%
Vasil74}%
\begin{APACrefauthors}%
{Vasil'Ev}, M\BPBI B.%
, {Vinogradov}, V\BPBI A.%
, {Vyshlov}, A\BPBI S.%
, {Ivanovskii}, O\BPBI G.%
, {Kolosov}, M\BPBI A.%
, {Savich}, N\BPBI A.%
\BDBL {}{Shtern}, D\BPBI Y.%
\end{APACrefauthors}%
\unskip\
\newblock
\APACrefYearMonthDay{1974}{{\APACmonth{01}}}{}.
\newblock
{\BBOQ}\APACrefatitle {{Radio Transparency of Circumlunar Space Using the
  Luna-19 Station}} {{Radio Transparency of Circumlunar Space Using the Luna-19
  Station}}.{\BBCQ}
\newblock
\APACjournalVolNumPages{Cosmic Research}{12}{}{102}.
\PrintBackRefs{\CurrentBib}

\bibitem [\protect \citeauthoryear {%
{Vocks, C.}%
\ \protect \BOthers {.}}{%
{Vocks, C.}%
\ \protect \BOthers {.}}{%
{\protect \APACyear {2018}}%
}]{%
Vocks18}
\APACinsertmetastar {%
Vocks18}%
\begin{APACrefauthors}%
{Vocks, C.}%
, {Mann, G.}%
, {Breitling, F.}%
, {Bisi, M. M.}%
, {Dabrowski, B.}%
, {Fallows, R.}%
\BDBL {}{Rucker, H.}%
\end{APACrefauthors}%
\unskip\
\newblock
\APACrefYearMonthDay{2018}{}{}.
\newblock
{\BBOQ}\APACrefatitle {LOFAR observations of the quiet solar corona} {Lofar
  observations of the quiet solar corona}.{\BBCQ}
\newblock
\APACjournalVolNumPages{A\&A}{614}{}{A54}.
\newblock
\begin{APACrefURL} \url{https://doi.org/10.1051/0004-6361/201630067}
  \end{APACrefURL}
\newblock
\begin{APACrefDOI} \doi{10.1051/0004-6361/201630067} \end{APACrefDOI}
\PrintBackRefs{\CurrentBib}

\bibitem [\protect \citeauthoryear {%
{Vyshlov}%
}{%
{Vyshlov}%
}{%
{\protect \APACyear {1976}}%
}]{%
Vyshlov76}
\APACinsertmetastar {%
Vyshlov76}%
\begin{APACrefauthors}%
{Vyshlov}, A\BPBI S.%
\end{APACrefauthors}%
\unskip\
\newblock
\APACrefYearMonthDay{1976}{}{}.
\newblock
{\BBOQ}\APACrefatitle {{Preliminary results of circumlunar plasma research by
  the Luna 22 spacecraft}} {{Preliminary results of circumlunar plasma research
  by the Luna 22 spacecraft}}.{\BBCQ}
\newblock
\BIn{} M\BPBI J.~{Rycroft}\ (\BED), \APACrefbtitle {Space research XVI} {Space
  research xvi}\ (\BPG~945-949).
\PrintBackRefs{\CurrentBib}

\bibitem [\protect \citeauthoryear {%
{Vyshlov}%
\ \BBA {} {Savich}%
}{%
{Vyshlov}%
\ \BBA {} {Savich}%
}{%
{\protect \APACyear {1979}}%
}]{%
VySav79}
\APACinsertmetastar {%
VySav79}%
\begin{APACrefauthors}%
{Vyshlov}, A\BPBI S.%
\BCBT {}\ \BBA {} {Savich}, N\BPBI A.%
\end{APACrefauthors}%
\unskip\
\newblock
\APACrefYearMonthDay{1979}{{\APACmonth{01}}}{}.
\newblock
{\BBOQ}\APACrefatitle {{Observations of radio source occultations by the moon
  and the nature of the plasma near the moon}} {{Observations of radio source
  occultations by the moon and the nature of the plasma near the moon}}.{\BBCQ}
\newblock
\APACjournalVolNumPages{Cosmic Research}{16}{}{551-556}.
\PrintBackRefs{\CurrentBib}

\bibitem [\protect \citeauthoryear {%
Woody%
}{%
Woody%
}{%
{\protect \APACyear {2001}}%
{\protect \APACexlab {{\protect \BCnt {1}}}}}]{%
WoodyALMA390}
\APACinsertmetastar {%
WoodyALMA390}%
\begin{APACrefauthors}%
Woody, D.%
\end{APACrefauthors}%
\unskip\
\newblock
\APACrefYearMonthDay{2001{\protect \BCnt {1}}}{}{}.
\newblock
{\BBOQ}\APACrefatitle {Radio Interferometer Array Point Spread Functions II.
  Evaluation and Optimization} {Radio interferometer array point spread
  functions ii. evaluation and optimization}.{\BBCQ}
\newblock
\APACjournalVolNumPages{ALMA Memo Series}{390}{}{}.
\PrintBackRefs{\CurrentBib}

\bibitem [\protect \citeauthoryear {%
Woody%
}{%
Woody%
}{%
{\protect \APACyear {2001}}%
{\protect \APACexlab {{\protect \BCnt {2}}}}}]{%
WoodyALMA389}
\APACinsertmetastar {%
WoodyALMA389}%
\begin{APACrefauthors}%
Woody, D.%
\end{APACrefauthors}%
\unskip\
\newblock
\APACrefYearMonthDay{2001{\protect \BCnt {2}}}{}{}.
\newblock
{\BBOQ}\APACrefatitle {Radio Interferometer Array Point Spread Functions I.
  Theory and Statistics} {Radio interferometer array point spread functions i.
  theory and statistics}.{\BBCQ}
\newblock
\APACjournalVolNumPages{ALMA Memo Series}{389}{}{}.
\PrintBackRefs{\CurrentBib}

\bibitem [\protect \citeauthoryear {%
{Wu}%
\ \BBA {} {Lee}%
}{%
{Wu}%
\ \BBA {} {Lee}%
}{%
{\protect \APACyear {1979}}%
}]{%
Wu79}
\APACinsertmetastar {%
Wu79}%
\begin{APACrefauthors}%
{Wu}, C\BPBI S.%
\BCBT {}\ \BBA {} {Lee}, L\BPBI C.%
\end{APACrefauthors}%
\unskip\
\newblock
\APACrefYearMonthDay{1979}{{\APACmonth{06}}}{}.
\newblock
{\BBOQ}\APACrefatitle {{A theory of the terrestrial kilometric radiation}} {{A
  theory of the terrestrial kilometric radiation}}.{\BBCQ}
\newblock
\APACjournalVolNumPages{The Astrophysical Journal}{230}{}{621-626}.
\newblock
\begin{APACrefDOI} \doi{10.1086/157120} \end{APACrefDOI}
\PrintBackRefs{\CurrentBib}

\bibitem [\protect \citeauthoryear {%
Zaslavsky%
, Meyer-Vernet%
, Hoang%
, Maksimovic%
\BCBL {}\ \BBA {} Bale%
}{%
Zaslavsky%
\ \protect \BOthers {.}}{%
{\protect \APACyear {2011}}%
}]{%
Zaslavsky11}
\APACinsertmetastar {%
Zaslavsky11}%
\begin{APACrefauthors}%
Zaslavsky, A.%
, Meyer-Vernet, N.%
, Hoang, S.%
, Maksimovic, M.%
\BCBL {}\ \BBA {} Bale, S\BPBI D.%
\end{APACrefauthors}%
\unskip\
\newblock
\APACrefYearMonthDay{2011}{}{}.
\newblock
{\BBOQ}\APACrefatitle {On the antenna calibration of space radio instruments
  using the galactic background: General formulas and application to
  STEREO/WAVES} {On the antenna calibration of space radio instruments using
  the galactic background: General formulas and application to
  stereo/waves}.{\BBCQ}
\newblock
\APACjournalVolNumPages{Radio Science}{46}{2}{}.
\newblock
\begin{APACrefURL}
  \url{https://agupubs.onlinelibrary.wiley.com/doi/abs/10.1029/2010RS004464}
  \end{APACrefURL}
\newblock
\begin{APACrefDOI} \doi{10.1029/2010RS004464} \end{APACrefDOI}
\PrintBackRefs{\CurrentBib}

\bibitem [\protect \citeauthoryear {%
Zyma%
, Girard%
, Landquist%
, Schaper%
\BCBL {}\ \BBA {} Vasko%
}{%
Zyma%
\ \protect \BOthers {.}}{%
{\protect \APACyear {2017}}%
}]{%
Zyma17}
\APACinsertmetastar {%
Zyma17}%
\begin{APACrefauthors}%
Zyma, K.%
, Girard, J\BPBI N.%
, Landquist, E.%
, Schaper, G.%
\BCBL {}\ \BBA {} Vasko, F\BPBI J.%
\end{APACrefauthors}%
\unskip\
\newblock
\APACrefYearMonthDay{2017}{}{}.
\newblock
{\BBOQ}\APACrefatitle {Formulating and solving a radio astronomy antenna
  connection problem as a generalized cable-trench problem: an empirical study}
  {Formulating and solving a radio astronomy antenna connection problem as a
  generalized cable-trench problem: an empirical study}.{\BBCQ}
\newblock
\APACjournalVolNumPages{International Transactions in Operational
  Research}{24}{5}{943-957}.
\newblock
\begin{APACrefURL}
  \url{https://onlinelibrary.wiley.com/doi/abs/10.1111/itor.12312}
  \end{APACrefURL}
\newblock
\begin{APACrefDOI} \doi{10.1111/itor.12312} \end{APACrefDOI}
\PrintBackRefs{\CurrentBib}

\end{thebibliography}

%
%
%
%
%

\end{document}